\input harvmac.tex

\newdimen\tableauside\tableauside=1.0ex
\newdimen\tableaurule\tableaurule=0.4pt
\newdimen\tableaustep
\def\phantomhrule#1{\hbox{\vbox to0pt{\hrule height\tableaurule width#1\vss}}}
\def\phantomvrule#1{\vbox{\hbox to0pt{\vrule width\tableaurule height#1\hss}}}
\def\sqr{\vbox{%
  \phantomhrule\tableaustep
  \hbox{\phantomvrule\tableaustep\kern\tableaustep\phantomvrule\tableaustep}%
  \hbox{\vbox{\phantomhrule\tableauside}\kern-\tableaurule}}}
\def\squares#1{\hbox{\count0=#1\noindent\loop\sqr
  \advance\count0 by-1 \ifnum\count0>0\repeat}}
\def\tableau#1{\vcenter{\offinterlineskip
  \tableaustep=\tableauside\advance\tableaustep by-\tableaurule
  \kern\normallineskip\hbox
    {\kern\normallineskip\vbox
      {\gettableau#1 0 }%
     \kern\normallineskip\kern\tableaurule}%
  \kern\normallineskip\kern\tableaurule}}
\def\gettableau#1 {\ifnum#1=0\let\next=\null\else
  \squares{#1}\let\next=\gettableau\fi\next}

\tableauside=1.0ex
\tableaurule=0.4pt

\lref\cs{E. Witten, ``Quantum field theory and the Jones polynomial," 
Commun. Math. Phys. {\bf 121} (1989) 351.}
\lref\gvone{R. Gopakumar and C. Vafa, ``Topological gravity 
as large $N$ topological gauge theory," hep-th/9802016, 
Adv. Theor. Math. Phys. {\bf 2} (1998) 413.}
\lref\gvtwo{R. Gopakumar and C. Vafa, ``On the gauge theory/geometry 
correspondence," hep-th/9811131.}
\lref\ov{H. Ooguri and C. Vafa, ``Knot invariants and topological 
strings," hep-th/9912123, Nucl. Phys. {\bf B 577} (2000) 419.} 
\lref\cmr{S. Cordes, G. Moore and S. Ramgoolam, ``Lectures on  
two-dimensional Yang-Mills theory, equivariant cohomology, and 
topological field theory," hep-th/9412210, Nucl. Phys. Proc. Suppl. {\bf 41} 
(1995) 184.}
\lref\gv{R. Gopakumar and C. Vafa, ``M-theory and topological 
strings, I," hep-th/9809187. }
\lref\gvm{R. Gopakumar and C. Vafa, ``M-theory and topological 
strings, II," hep-th/9812127. }
\lref\kkv{S. Katz, A. Klemm and C. Vafa, ``M-theory, topological strings, and 
spinning black-holes," hep-th/9910181.}
\lref\dz{J.M. Drouffe and J.B. Zuber, ``Strong-coupling and mean 
field methods in lattice gauge theories," Phys. Rep. {\bf 102} (1983) 1.}
\lref\llr{J.M.F. Labastida, P.M. Llatas and A.V. Ramallo, ``Knot operators 
in Chern-Simons theory," Nucl. Phys. {\bf B 348} (1991) 651.}
\lref\lm{J.M.F. Labastida and M. Mari\~no, ``The HOMFLY polynomial for 
torus links from Chern-Simons gauge theory," hep-th/9402093, Int. J. Mod.
 Phys. {\bf A 10} (1995) 1045.} 
\lref\ilr{J.M. Isidro, J.M.F. Labastida, and A.V. Ramallo, 
``Polynomials for torus links from Chern-Simons gauge theories," 
hep-th/9210124, Nucl. Phys. {\bf B 398} (1993) 187.}
\lref\fh{W. Fulton and J. Harris, {\it Representation theory. A first course},
Springer-Verlag, 1991.}
\lref\jonesann{V.F.R. Jones, ``Hecke algebras representations of braid 
groups and link polynomials," Ann. of Math. {\bf 126} (1987) 335. }
\lref\guada{E. Guadagnini, ``The universal link polynomial," 
Int. J. Mod. Phys. {\bf A 7} (1992) 877;  {\it The link invariants of the 
Chern-Simons field theory,} Walter de Gruyter, 1993.}
\lref\lick{W.B.R. Lickorish, {\it An introduction to knot theory}, 
Springer-Verlag, 1998.}
\lref\awrep{M. Wadati, T. Deguchi and Y. Akutsu, ``Exactly solvable models 
and knot theory," Phys. Rep. {\bf 180} (1989) 247.}
\lref\ofer{O. Aharony, S. Gubser, J. Maldacena, H. Ooguri and Y. Oz, 
``Large $N$ field theories, string theory and gravity," hep-th/9905111, 
Phys. Rep. {\bf 323} (2000) 183.}
\lref\lp{J.M.F. Labastida and E. P\'erez, ``A relation between the 
Kauffman and the HOMFLY polynomial for torus knots," q-alg/9507031, 
J. Math. Phys. {\bf 37} (1996) 2013.}
\lref\homfly{P. Freyd, D. Yetter, J. Hoste, W.B.R. Lickorish, K. Millet and 
A. Ocneanu, ``A new polynomial invariant of knots and links," Bull. Amer. 
Math. Soc. {\bf 12} (1985) 239.}
\lref\d{M.R. Douglas, ``Conformal field theory techniques in large $N$ 
Yang-Mills theory," hep-th/9311130.}
\lref\quanta{S.~Elitzur, G.~Moore, A.~Schwimmer and N.~Seiberg,
``Remarks on the canonical quantization of the Chern-Simons-Witten theory,''
Nucl.\ Phys.\  {\bf B 326} (1989) 108. J.M.F. Labastida 
and A.V. Ramallo, ``Operator formalism for Chern-Simons theories,''
Phys.\ Lett.\  {\bf B 227}(1989) 92; ``Chern-Simons theory and conformal 
blocks,'' Phys.\ Lett.\  {\bf B 228} (1989) 214. S. Axelrod, S. Della Pietra, 
and E. Witten, ``Geometric quantization of Chern-Simons gauge theory,'' 
J. Diff. Geom. {\bf 33} (1991) 787.} 
\lref\al{M. \'Alvarez and J.M.F. Labastida, ``Vassiliev invariants for torus 
knots,'' q-alg/9506009, J. Knot Theory Ramifications {\bf 5} (1996) 779.}
\lref\rama{P. Ramadevi, T.R. Govindarajan and R.K. Kaul, ``Three-dimensional 
Chern-Simons theory as a theory of knots and links (III). 
Compact semi-simple group,'' Nucl.\ Phys.\ {\bf
B 402} (1993) 548.}
\lref\lpp{J.M.F. Labastida and E. P\'erez, ``Gauge-invariant 
operators for singular knots in
Chern-Simons gauge theory," hep-th/9712139, Nucl.\ Phys.\ {\bf
B 527} (1998) 499.}
\lref\witop{E. Witten, ``Chern-Simons gauge theory as 
a string theory,'' hep-th/9207094, in {\it The Floer memorial volume}, 
H. Hofer, C.H. Taubes, A. Weinstein and E. Zehner, eds., 
Birkh\"auser 1995, p. 637.}  
\lref\peri{V. Periwal, ``Topological closed-string theory 
interpretation of Chern-Simons theory,'' hep-th/9305112, Phys. Rev. Lett. 
{\bf 71} (1993) 1295.}
\lref\douglasn{M.R. Douglas, ``Chern-Simons-Witten theory as a topological 
Fermi liquid,'' hep-th/9403119.}
\lref\vass{V.A. Vassiliev, ``Cohomology of knot spaces", {\it Theory
of singularities and its applications}, {\sl Advances in Soviet
Mathematics}, vol. 1, {\sl Americam Math. Soc.}, Providence, RI, 1990,
23-69.}
\lref\simon{S. Willerton, ``On universal Vassiliev invariants, cabling, 
and  torus knots", University of Melbourne preprint  (1998).}
\lref\awada{M.A. Awada, ``The exact equivalence of Chern-Simons 
theory with fermionic string theory,'' Phys. Lett. {\bf B 221} (1989) 21.}
 

\Title{\vbox{\baselineskip12pt
\hbox{US-FT-7/00}
\hbox{RUNHETC-2000-14}
\hbox{hep-th/0004196}
}}
{\vbox{\centerline{Polynomial Invariants for Torus Knots}
\centerline{ }
\centerline{and Topological Strings}}
}
\centerline{J.M.F. Labastida$^{a}$ and Marcos Mari\~no$^{b}$}

\bigskip
\medskip
{\vbox{\centerline{$^{a}$ \sl Departamento de F\'\i sica de Part\'\i culas}
\centerline{\sl Universidade de Santiago de Compostela}
\vskip2pt
\centerline{\sl E-15706 Santiago de Compostela, Spain}}
\centerline{ \it labasti@fpaxp1.usc.es}

\bigskip
\medskip
{\vbox{\centerline{$^{b}$ \sl New High Energy Theory Center}
 \centerline{\sl Rutgers University}
\vskip2pt
\centerline{\sl Piscataway, NJ 08855, USA }}
\centerline{ \it marcosm@physics.rutgers.edu }

\bigskip
\bigskip
\noindent

We make a precision test  of a recently proposed conjecture relating Chern-Simons
gauge theory  to topological string theory on the resolution of the conifold. 
First, we develop a systematic procedure to extract string amplitudes  from
vacuum expectation values (vevs) of Wilson loops in Chern-Simons gauge theory,
and  then we evaluate these vevs in arbitrary irreducible representations of
$SU(N)$ for  torus knots. We find complete agreement with the predictions derived
from  the target space interpretation  of the string amplitudes. We also show
that the structure of the free energy of topological 
open string theory gives
further constraints  on the Chern-Simons vevs. 
Our work provides strong evidence
towards an interpretation of knot polynomial 
invariants as generating functions
associated to enumerative problems.

\bigskip

\Date{April, 2000}

\listtoc \writetoc
\newsec{Introduction}

Ever since the Jones polynomial and its generalizations were 
discovered \jonesann,
knot theorists have been searching for an interpretation of the 
integers entering
these polynomials. Though it seems rather natural to regard 
these polynomials as
generating functions associated to enumerative problems, no much 
progress has been
achieved in this direction. One of the main goals of this paper 
is to point out
that the situation has changed dramatically after the recent 
work by Ooguri and
Vafa in \ov. Based on their results, we will 
provide strong evidence to affirm
that from the ordinary polynomial invariants 
associated to arbitrary irreducible
representations of the group $SU(N)$ one can 
construct new ones whose integer
coefficients can be interpreted as the solutions to specific 
enumerative problems in
the context of string theory. Thus, in what regards to a picture of polynomial
invariants as generating functions, these new polynomials are more fundamental
than the ordinary ones. At the heart of this development is Chern-Simons gauge
theory \cs\ and the relationship between large $N$ gauge theories and  gravity in
the light of the AdS/CFT correspondence (see
\ofer\ for a review).

 The proposal of \gvone\gv\gvtwo\ov, which can be regarded as a simpler 
version of the AdS/CFT correspondence, relates Chern-Simons
gauge theory  on ${\bf S}^3$ to topological string theory whose target is the 
resolution of the conifold\foot{The relation between  Chern-Simons gauge theory
and string theory has been addressed also in
\awada\peri\douglasn. The connection between Chern-Simons and  topological open string
theory was discovered by Witten in \witop. }. 
 This  proposal is very interesting from  the point of view of
knot theory and three-manifold topology, since  it reformulates the invariants
obtained in the context of Chern-Simons gauge theory  in terms of invariants
associated to topological strings and  related to the counting 
of BPS states. In particular, in \ov\ a generating  function of
vacuum expectation values (vevs) of Wilson loops was expressed  in terms of
certain integers counting the number of D2 branes ending  on D4 branes. This
reformulation makes some predictions  about the structure of Chern-Simons vevs
and provides an interpretation for the integer coefficients of some related
polynomial invariants. It was verified in \ov\ that these  predictions were true
in the simple case of the unknot. 

One purpose of this paper is to  make a precision test of the proposal of \ov\
for a wide class of nontrivial  knots.  As a preliminary step, we present a
systematic  procedure to extract from the vevs of  Wilson loops a series of
polynomials arising naturally in the  context of topological strings. This is the
content of equation (2.19)  below. These polynomials, that we denote\foot{Though
we will refer to the  $f_R$ as polynomials, they are not. They are polynomials up
to a common factor as stated in (2.8) and (4.22).} by
$f_R$, are labeled by the irreducible representations, $R$, of $SU(N)$, and
according  to the conjecture in \ov\ they have a very precise structure dictated
by the BPS content  of the ``dual" theory. We then test this conjecture with
actual  computations in Chern-Simons gauge theory.

The technical challenge associated to the  conjecture in \ov, on the Chern-Simons
side,
 is that it involves vevs of Wilson loops in  arbitrary irreducible
representations of
$SU(N)$, with $N$ generic.  For the fundamental representation,  the vevs are
related (up to a normalization) to the HOMFLY polynomial 
\homfly. Not much is known about these vevs  for other irreducible
representations, except in the case of $SU(2)$, where they are  related to the
Akutsu-Wadati polynomials (for a review, see \awrep).  There are also some sample
computations for a few knots in \rama,   for representations of
$SU(N)$ with only one row in their Young diagram\foot{There are also  a few
computations in \guada\ for the gauge group $SU(3)$.}. However, in  the case of
torus knots,  one can compute these vevs  using the formalism of knot operators
introduced in \llr.  Knot operators were used in \llr\ilr\lm\lp\  to compute the
vevs of Wilson loops for torus knots and links  in the fundamental
representations of $SU(N)$ and $SO(N)$, and in arbitrary  irreducible
representations of
$SU(2)$. The computation of vevs for torus knots  in arbitrary irreducible
representations of
$SU(N)$, as needed to test the conjecture  of \ov, is technically difficult, but
fortunately many of the  intermediate results were already obtained in \lm. This
leads to a  general formula for these vevs, which can be found in (3.29)  below.
Despite its intimidating aspect, it is not difficult  to implement it in a
computer routine to obtain  the vevs for any torus knot. The equation (3.29) is
of course a result  interesting in its own, and we hope that it will be helpful
in exploring the  generalizations of the HOMFLY polynomial to arbitrary
irreducible representations  of $SU(N)$. 

Using the general formula (3.29), we will test 
the conjecture presented in \ov\ for some nontrivial knots. We will find that, in
all the  examples that we have checked, the polynomials $f_R$ have in fact 
the structure predicted by \ov. This is a highly nontrivial fact from the 
point of view of Chern-Simons gauge theory, and we regard it as a strong 
evidence for the duality advocated in \gvone\gv\gvtwo\ov. 

There are in fact two different predictions in \ov, which are 
in a sense complementary. The first one predicts 
the structure of the polynomials $f_R$, it is based on a target 
space interpretation, and it is nonperturbative. The second one 
is perturbative 
and it is based on the worldsheet interpretation of the Chern-Simons vevs 
presented in \witop. These two predictions are related in a very interesting 
way. More precisely, it turns out that the perturbative structure of the free 
energy of the open string gives some ``sum rules'' on the integers 
that count BPS configurations. We have also found complete agreement with the 
perturbative prediction in all the examples that 
we have checked.   

The paper is organized as follows: in section 2, we describe the conjecture 
presented in ref. \ov, which expresses a generating functional of Chern-Simons
gauge theory  in terms of certain polynomials $f_R$. We extract from the
conjecture a  ``master equation" which allows us to obtain these 
functions from 
usual vevs in Chern-Simons gauge  theory through a recursive procedure. In
section 3, we obtain a  general formula for the vevs of torus knots 
in arbitrary
irreducible representations of 
$SU(N)$. This section contains the arguments leading to formula (3.29), which are
independent of the rest of the paper. It could be skipped in a first reading. In
section 4, we use formula (3.29) to obtain some of the
polynomials
$f_R$, taking  as an example the right-handed trefoil knot. The results are in
full  agreement with the conjecture of \ov.  In section 5 
we show that the perturbative point of view gives some nontrivial constraints 
among the integer invariants that appear in the polynomials $f_R$, and 
we also show that the connected vevs of Chern-Simons have the structure 
dictated by these constraints. Finally, in section 6, we
conclude with some comments and open problems.  An appendix collects the
expressions of $f_R$ for the right-handed trefoil knot for all irreducible
representations of $SU(N)$ whose associated Young Tableaux contains four
boxes.    

\newsec{Extracting string amplitudes from Chern-Simons gauge theory}

We first recall some basic aspects of Chern-Simons gauge theory, mainly to 
fix our notation. Chern-Simons gauge theory is 
a topological gauge theory whose action is, 
\eqn\csaction{
S={k \over 4\pi} \int_M {\rm Tr} \Bigl( A\wedge d A + {2 \over 3} A
\wedge A \wedge A \Bigr),} 
where $A$ is a gauge connection on some vector bundle over a
 three-manifold $M$, and $k$ is the coupling constant. 
From the holonomy of 
the gauge field around a closed loop $\gamma$ in $M$,
\eqn\holo{
U={\rm P}\,\exp\, \oint_{\gamma} A,}
one can construct a natural class of topological observables, the 
gauge-invariant Wilson loop operators, which are given by
\eqn\wilso{
W^{\gamma}_R(A)={\rm Tr}_R \, U,}
where $R$ denotes an irreducible representation of $SU(N)$.
Some of the standard topological invariants that have been considered in the
context of Chern-Simons gauge theory are vevs of products of these operators:
\eqn\correlat{
\langle W^{\gamma_1}_{R_1}\cdots W^{\gamma_n}_{R_n}\rangle =
{1\over Z(M)}\int [{\cal D} A] \Bigl( \prod_{i=1}^n W_{R_i}^{\gamma_i} 
\Bigr) {\rm e}^{iS},}
where $Z(M)$ is the partition function of the theory. In this paper we will
consider an enlarged set of operators which, to our knowledge, has not been
studied from a  Chern-Simons gauge theory point of view for non-trivial knots.
These operators involve, besides the standard Wilson loops and their products,
additional products with traces of powers of the holonomy \holo. We will 
compute
their vevs for the case of torus knots. In the process we
will derive a formula for the vevs of Wilson loops in arbitrary 
irreducible representations of the gauge group $SU(N)$. The resulting vevs  will
be expressed in terms of the variables\foot{A word of caution about  notation: in
\lm, the variable $\lambda$ is denoted $t^{N-1}$. Also, in order  to compare to
\ov, notice that our $t$ is their
$\exp(i\lambda)$, and  our $\lambda$ is their $\exp t$.}, 
\eqn\variar{
t=\exp \Bigl[ {2 \pi i \over k +N} \Bigr],\,\,\,\,\,\ \lambda =t^N.} 

In order to make a precise test of the conjecture presented in ref. \ov, we
will consider the vev of the operator, 
\eqn\gen{
Z(U,V)=\exp\bigl[ \sum_{n=1}^\infty {1 \over n} {\rm Tr}\, U^n\, 
{\rm Tr }\, V^n\bigr],}
where $U$ is the holonomy of the Chern-Simons $SU(N)$ gauge field \holo, 
and $V$ is an $SU(M)$ matrix that can be regarded as a source term. In this
operator the trace is taken over the fundamental representation. In what follows,
when no representation is indicated in a trace, it should be understood that it
must be taken in the fundamental representation.

The main conjecture of \ov\ has two parts. First, it states that the vev
of \gen\ can be written as,
\eqn\conj{
\langle Z(U,V) \rangle =\exp\Big(\sum_{n=1}^\infty \sum_{R} f_R (t^n, \lambda^n) {\rm Tr}_R 
{V^n\over n} \Big),}
where the sum over $R$ is a sum over irreducible representations of 
$SU(M)$. Second, it predicts the following structure for the functions
$f_R(t, \lambda)$:
\eqn\ovconj{
f_R(t, \lambda)=\sum_{s,Q}{ N_{R,Q,s}\over t^{{1\over 2}}-t^{-{1\over 2}}}
 \lambda^Q t^s ,}
where $N_{R,Q,s}$ are integer numbers, and the $Q$ and $s$ are, in general, 
half-integers (however, for a given $f_R$, the 
$Q$ differ by integer numbers).  In writing \conj, and to be able to 
compare to the results in Chern-Simons gauge  theory, we have performed 
an analytic continuation, as suggested in \ov. The prediction \ovconj\ 
is based on the duality between Chern-Simons theory and topological string 
theory. As explained in \ov, given a knot ${\cal K}$ in 
${\bf S}^3$ one constructs a Lagrangian submanifold ${\cal C}_{\cal K}$ 
in the noncompact Calabi-Yau ${\cal O}(-1) + {\cal O}(-1) \rightarrow 
{\bf S}^2$ (the resolution of the conifold). The integers $N_{R,Q,s}$ 
count, very roughly, holomorphic maps from Riemann surfaces with boundaries 
to the 
Calabi-Yau, in such a way that the boundaries are mapped to  
${\cal C}_{\cal K}$. A more precise understanding of the integers 
$N_{R,Q,s}$ is given by the target space 
interpretation of the string amplitudes. In this interpretation, one 
reformulates the counting problem in terms of D-branes.    
One considers 
configurations of D2 branes ending on ${\cal C}_{\cal K}$, in the 
presence of $M$ D4 branes wrapping ${\cal C}_{\cal K}$ and filling an 
${\bf R} ^2$ in the uncompactified spacetime. The D2 branes are BPS particles 
from the two-dimensional point of view. These particles are characterized 
by their magnetic charge, their bulk D2 brane charge, and their spin, which
 correspond, respectively, to $R$, $Q$ and $s$ in 
\ovconj. The integer $N_{R,Q,s}$ counts the number of BPS states with 
these quantum numbers. We then see that the conjecture of \ov\ makes 
a remarkable connection between knot invariants and an enumerative problem 
in the context of symplectic and algebraic geometry, and that the 
polynomials $f_R$ can be regarded as counting functions for this enumerative 
problem. 

In this section we will prove the
first part of the conjecture. It follows from simple group theoretical
arguments. Thus, it will be established that the vevs
 of Wilson loops in arbitrary irreducible representations 
of the gauge group 
can be encoded in the functions $f_R(t, \lambda)$. This also gives 
a concrete procedure to compute these functions from Chern-Simons vevs, 
and using this procedure we will present a highly nontrivial evidence 
for \ovconj\ in the case of torus knots.

Our starting point is the construction of a set of linear equations for the
functions $f_R(t, \lambda)$ in terms of vevs of
standard Wilson loops in arbitrary irreducible representations. 
To carry this out, 
it is convenient to use the following basis of class functions (see, for 
example, 
\dz\cmr\d). Take a vector $\vec k$ with an infinite number of entries, 
almost all 
zero, and whose nonzero entries are positive integers. Given such a vector, 
we define:
\eqn\long{
\ell=\sum j \,k_j,\,\,\,\,\,\ |\vec k| =\sum k_j.} 
We can associate to any vector $\vec k$ a conjugacy class $C(\vec k)$ of the  
permutation group  $S_\ell$. This class has $k_1$ cycles of 
length 1, $k_2$ cycles of length 2, and so on. The number of elements 
of the permutation group in such a 
class is given by \fh\
\eqn\conjug{
|C(\vec k)|= {\ell! \over \prod k_j! \prod j^{k_j}}.}
Equivalently, the vectors $\vec k$ with $\sum_j\, j k_j=\ell$ 
are in one-to-one 
correspondence with the partitions of $\ell$. Given an ordered $h$-uple 
of positive integers $(n_1, \cdots, n_h)$, we can map it to a 
vector $\vec k$ by putting $k_i$ equal to the number of $i$'s in the 
$h$-uple. Notice that $h=|\vec k|$, and that there are $h!/\prod k_j!$ 
different $h$-uples giving the same vector $\vec k$. 

We now introduce the following basis in the space of class functions, 
labeled by the vectors $\vec k$:
\eqn\basis{
\Upsilon_{\vec k}(U)= \prod_{j=1}^\infty \Big( {\rm Tr}\,U^j \Big)^{k_j}.}
It is
 easy to see that:
\eqn\rewr{
Z(U,V)=1+\sum_{\vec k} {|C(\vec k)| \over \ell!}
 \Upsilon_{\vec k} (U) \Upsilon_{\vec k}(V),}   
since we are assuming $\ell>0$.

Let's now consider the expansion of the exponent in \conj\ in terms of the 
basis \basis. We first recall Frobenius formula to express  traces in an 
arbitrary irreducible representation of $SU(M)$ in terms of the elements of the
basis \basis\ referred to this group:
\eqn\frob{
{\rm Tr}_R (V)=\sum_{\vec k} {|C(\vec k)| \over \ell!} \chi_R ( C(\vec k)) 
\Upsilon_{\vec k} (V).}
In this formula, the irreducible representation $R$ can be associated to a Young
diagram   in the standard way. The sum is then over conjugacy classes with 
$\ell$ equal to  the number of boxes in the diagram. 

To analyze the expansion in \conj, we have to write  
${\rm Tr}_R V^n $ in terms of the basis \basis. To do this it is convenient to
define the following  vector $\vec k_{1/n}$. Fix a vector $\vec k$, and 
consider all the positive integers that satisfy the following condition:
$n|j$ for every $j$ with $k_j \not=0$. Notice that $n=1$ always 
satisfies this condition. When this happens, we will say that ``$n$ divides 
$\vec k$,'' and we will denote this as $n|\vec k$.  
We can then define the vector $\vec k_{1/n}$ whose components are:
\eqn\shifted{
(\vec k_{1/n})_i=k_{ni}.}
The vectors which satisfy the above condition and are
 ``divisible by $n$'' have the structure 
$(0,\dots, k_n, 0,\dots,0, k_{2n},\dots)$, and the vector 
$\vec k_{1/n}$ is then
given by $(k_n,k_{2n},\dots)$. It is a simple combinatorial 
exercise to prove 
that the exponent in \conj\ is given by:
\eqn\enes{
\sum_{\vec k}{|C(\vec k)|\over \ell!} \sum_{n|\vec k} n^{|\vec k| -1} 
 \sum_R \chi_R (C(\vec k_{1/n}))f_R(t^n, \lambda^n)\Upsilon_{\vec k}(V) .}
In this equation, the third 
sum is over representations 
of $S_{\ell}$. 
We now define a generalization of the cumulant expansion for the vevs 
we are 
considering. First, associate to any $\vec k$ the polynomial $p_{\vec k}(x)=
 \prod_j x_j^{k_j}$ 
in the variables $x_1, x_2, \dots$. We then define the
 ``connected" coefficients 
$a_{\vec k}^{(c)}$ as follows: 
\eqn\cumul{
\log\Bigg(1+ \sum_{\vec k} {|C(\vec k)| \over \ell!} a_{\vec k}p_{\vec k}(x) \Bigg)= 
\sum_{\vec k} {|C(\vec k)| \over \ell!} a^{(c)}_{\vec k}p_{\vec k}(x).}
One has, for example: 
\eqn\cumulas{
\eqalign{
a^{(c)}_{(2,0,\dots)}&=a_{(2,0,\dots)}-a^2_{(1,0,\dots)},\cr
a^{(c)}_{(1,1,0,\dots)}&=a_{(1,1,0,\dots)}- a_{(1,0,\dots)}a_{(0,1,0,\dots)},\cr
a^{(c)}_{(0,\dots,0,1,0,\dots)}&= a_{(0,\dots,0,1,0,\dots)},\cr}}
and so on. For vectors of the form $(n,0,\dots)$, this is just the cumulant 
expansion. 

Define now the vevs:
\eqn\vevs{
G_{\vec k}(U) = \langle \Upsilon_{\vec k} (U) \rangle.} 
Using \gen\rewr\cumul\ and \vevs, we find:
\eqn\lojari{
\log \, \langle Z(U,V) \rangle =\sum_{\vec k} {|C(\vec k)| \over \ell!} 
G_{\vec k}^{(c)}(U) \Upsilon_{\vec k}(V).}
Since the $\Upsilon_{\vec k}(V)$ are a basis in the space of class 
functions, we find that the equation
\conj\  can be written as 
\eqn\master{
G_{\vec k}^{(c)}(U)= \sum_{n|\vec k} n^{|\vec k| -1}
 \sum_R \chi_R (C(\vec k_{1/n}))f_R(t^n, \lambda^n).}
This is our ``master equation". It allows us to obtain the 
functions $f_R(t,\lambda)$ once we compute the vevs that appear on
the left hand side. The way to do that is to consider all vectors $\vec k$ with 
a fixed $\ell$, where $\ell$  will be considered as the ``order" of the expansion. 
The number of these vectors is the number of partitions of $\ell$,
 $p(\ell)$. At every 
order there are then $p(\ell)$ vevs $G_{\vec k}^{(c)}(U)$ and also $p(\ell)$ 
representations $R$ of $S_{\ell}$. The relations \master\ provide
$p(\ell)$ equations with $p(\ell)$ unknowns, the functions 
$f_R(t,\lambda)$. The data to solve the equations are the vevs \vevs\  
and the $f_{R'}(t,\lambda)$ 
for representations with $\ell'<\ell$ boxes. The procedure to find 
the polynomials is then recursive, and the structure one finds 
is very similar, in fact, 
to the recursive procedure which determines the integer invariants introduced 
in \gvm, as it is explained in \kkv.

The above system of linear equations has a unique solution. This follows from
the fact that the  associated matrix, $|C(\vec k)| \chi_R( C(\vec k))$,
is invertible  due to orthonormality of the characters. Thus, the first part of
the conjecture, eq. \conj, is proved. 

There are two cases of the above expression which 
are particularly interesting. The first one is for  
$\vec k =(\ell,0,\dots)$. In this case, the corresponding conjugacy 
class in $S_\ell $ is the identity,  
and one finds
\eqn\firstex{
\langle ({\rm Tr} \,U)^\ell \rangle^{(c)} = \sum_R ({\rm dim}\, R)
 f_R (t, \lambda),}
where the sum is over the representations $R$ of $S_\ell$. 
The left hand side is the 
usual connected vev. The second example corresponds to the vector 
$\vec k =(0, \dots,0,1,0,\dots)$, where the nonzero entry is in the 
$\ell$-th position. 
In this case, we have to sum in \master\ over all the 
divisors of $\ell$, that we will 
denote by $n$. The vector $\vec k_{1/n}$ is then 
$(0,\dots, 0,1,0,\dots)$, where 
the nonzero entry is in the $\ell/n$-th position. The characters 
$\chi (C(\vec k_{1/n}))$ are different from zero only for 
the hook representations, 
{\it i.e.}, those corresponding to Young diagrams with $(\ell/n)-s$ 
boxes in the 
first row, and one box in the remaining ones, for example,
\tableauside=1.5ex
\tableaurule=0.4pt
\eqn\hook{
\tableau{6 1 1 1 1}
}
\tableauside=1.0ex
\tableaurule=0.4pt
The character is then $(-1)^s$ (see \fh, 4.16). 
The formula \master\ reads in this case: 
\eqn\secex{
\langle {\rm Tr}\, U^\ell \rangle = \sum_{n|\ell}\sum_{s=0}^{\ell/n-1} (-1)^s 
f_{{\rm hook},s }(t^n, \lambda^n).}

\newsec{Polynomial invariants for torus knots in arbitrary irreducible
representations  of $SU(N)$}

In order to obtain the functions $f_R$ from the master equation \master\ one
needs to compute the connected functions $G_{\vec k}^{(c)}(U)$. After using
\vevs\ and the inverse of Frobenius formula \frob, it turns out that these involve
the computation of vevs of Wilson loops in arbitrary irreducible representations
of $SU(N)$. As stated in the introduction these vevs are known only for some
particular cases. In order to have a good testing ground of the conjecture \conj\
it would be desirable to have a formula for these vevs  valid for any
representation, at least for some particular class of knots. The goal of this
section is to derive such a formula for torus knots. The result is contained in
eq. (3.29) below. The arguments leading to it are independent of the rest of the
paper and thus this section could be skipped in a first reading. The techniques
used to obtain the formula (3.29) are based on the application of the operator
formalism to Chern-Simons gauge theory \quanta, that in the case of torus knots
leads to the useful concept of knot operators \llr.

\subsec{Knot operators}
Knot operators for torus knots were introduced in \llr. They allow the 
computation of vevs of Wilson loops corresponding to this type of knots for
arbitrary irreducible representations of
 the gauge 
group. The first piece of data we need to introduce these operators is 
the Hilbert
space of Chern-Simons gauge theory on a torus \quanta. This space has 
an orthonormal basis 
$| p\rangle$ labeled 
by weights  $p$ in the fundamental chamber of the weight lattice 
of $SU(N)$, ${\cal F}_l$, where $l=k+N$. We take as representatives of $p$ the ones of the form $ p=\sum_i p_i  \lambda_i$, 
where $ \lambda_i$, $i=1, \cdots, N-1$,  
are the fundamental weights, $p_i>0$ and $\sum_i p_i <l$. The vacuum is 
the state $| \rho \rangle$, where $\rho$ is the 
Weyl vector ({\it i.e.}, 
the sum of all the fundamental weights). 

Torus knots are labeled by two coprime integers $(n,m)$. They correspond 
to winding numbers around the two non-contractible classes of cycles, $A$ and $B$
on the torus.  Let $m$ be the number of times that the torus knot winds
around the axis of  the torus, and let $\Lambda$ be the highest weight of an
irreducible  representation. Then, the Wilson loop corresponding to that torus knot
is  represented by the following operator:
\eqn\knotop{
W_{\Lambda}^{(n,m)}|p\rangle = \sum_{\mu \in M_{\Lambda}}\exp \biggl[ 
-i\pi \mu ^2 {nm \over k+N} - 2\pi i {m \over k+N}  p \cdot  \mu 
\biggr] | p + n \mu\rangle .}
In this equation, $M_{\Lambda}$ is the set of weights corresponding to the 
irreducible representation $\Lambda$. 

To compute the vev of the Wilson loop around a torus knot in ${\bf S}^3$, one 
proceeds as follows: first of all, one makes a Heegard splitting of 
 ${\bf S}^3$ into two solid tori.  Then, one puts the torus knot on the 
surface of one of the solid tori by acting with the knot operator \knotop\ 
on the vacuum. Finally, one glues together the tori by performing an 
$S$-transformation. There is an extra subtlety related to the 
framing dependence in Chern-Simons gauge theory, since the vev computed 
in this way has to be corrected with a phase.  
In the standard framing the vev of the Wilson loop is given by:
\eqn\vevop{
\langle W_{ \Lambda}^{(n,m)}\rangle =
{\rm e}^{2 \pi i nm h_{ \rho + \Lambda}}
{\langle \rho | S W_{\Lambda}^{(n,m)}
| \rho\rangle \over \langle  \rho | S| \rho\rangle},}
where,
\eqn\conf{
h_{ p}={p^2 - \rho^2 \over 2(k+N)},} 
is the conformal weight of the primary fields in the
associated WZW model at level $k$. 

\subsec{Vacuum expectation values from knot operators}
Our next task is to provide a more precise expression for the vev \vevop. When
acting  with the knot operator \knotop\ on the vacuum, we get the set of  weights
 $\rho + n \mu$, where $ \mu \in M_{ \Lambda}$. These 
weights will have representatives in the fundamental chamber, which can be  
obtained by a series of Weyl reflections. If the representative
 has a vanishing 
component, then the corresponding state in the Hilbert space is zero due to 
antisymmetry of the wave function under Weyl reflections. The set 
of weights that have a nonzero representative in ${\cal F}_l$ 
 will be denoted by $ {\cal M} (n, \Lambda)$, 
and it depends on the irreducible representation with
 highest weight $\Lambda$, and on the integer number
 $n$. The representative 
of $ \rho + n \mu$ in $ M (n, \Lambda)$ 
will be denoted by $\rho+\mu_n$. The matrix elements 
of $S$ have the explicit expression,
\eqn\smat{
S_{ p,p'}=c(N,k) \sum_{w\in {\cal W}} \epsilon (w) \exp 
\biggl[ -{2\pi i  p \cdot w( p') \over k+N} \biggr],}
where $c(N,k)$ is a constant depending only on $N$ and $k$, and the sum 
is over the Weyl group of $SU(N)$, ${\cal W}$. Using this, the vev \vevop\ can be 
written as:
\eqn\vevch{    
 {\rm e}^{2\pi i nm h_{ \rho +  \Lambda} }
\sum_{ \mu \in {\cal M}(n,  \Lambda)}
\exp \biggl[ 
-i\pi\mu ^2 {nm \over k+N} - 2\pi i {m \over k+N}  p \cdot  \mu 
\biggr] {\rm ch}_{ \mu_n}\Bigl[ -{ 2\pi i \over k+N} \rho\Bigr].}
In this expression, we have used the Weyl formula for the character:
\eqn\weyl{
{\rm ch}_{ \Lambda}( a)={\sum _{w \in {\cal W}} \epsilon (w) {\rm e}^
{ w( \Lambda +  \rho)\cdot  a} \over
\sum _{w \in {\cal W}} \epsilon (w) {\rm e}^
{ w( \rho)\cdot a}}.}
Notice that, since the representatives $\mu_n$ live in ${\cal F}_l$, they can 
be considered as highest weights for a representation, hence the above 
expression \vevch\ makes sense. 

In practice, the main problem to compute this vevs 
explicitly is to find the nonzero representatives of the weights that 
appear in \knotop, and to find an expression for the characters in 
\vevch. Fortunately, this has been done in \lm\ in a slightly different 
context. In that paper, these problems were solved for all the weights in the 
product representation $V^{\otimes s}$, where $V$ is the fundamental 
representation of $SU(N)$ and $s$ is any integer. Since all the 
representations of $SU(N)$ that correspond to Young diagrams 
with $s$ boxes are in fact contained
 in the reducible tensor product $V^{\otimes s}$, we only have to combine 
the results of \lm\ with some simple group theory. This will give an explicit 
expression for the vev value of Wilson loops for torus knots in 
{\it arbitrary} 
representations of $SU(N)$. 

\subsec{Group theory}
To obtain the expression for the vev of the Wilson loop, we need the 
weight space associated to arbitrary representations of $SU(N)$. It is 
very convenient to regard this space as a subspace of the weight space 
associated to the reducible representation $V^{\otimes s}$.   
Let's denote by $ \mu_i$, $i=1, \dots, N$ the weights of the fundamental
representation of $SU(N)$. Any weight in $V^{\otimes s}$ will have the form
\eqn\wpro{
k_1  \mu_{i_1} + \cdots + k_r \mu_{i_r}, \,\, 
1\le i_1< \dots <i_r\le N,}
where $(k_{\lambda})=(k_1, \dots, k_r)$ is an {\it ordered} 
partition of $s$, {\it i.e.} an $r$-tuple that sums up to $s$. 
The $k_{\lambda}$ will be taken 
as strictly positive integers, therefore $1\le r\le s$. 
The corresponding 
unordered partition will be simply denoted by $k$. Unordered partitions for 
$SU(N)$ will be written as $N$-tuples with nonincreasing components, as in 
\fh. The  
set of weights \wpro, for a fixed $(k_{\lambda})$, will be 
denoted by $M_{k_{\lambda}}$. 

Consider now a 
irreducible representation $R$ of $SU(N)$, associated to the highest weight   
\eqn\hw{
\Lambda =\sum_{i=1}^{N-1}a_i \lambda_i.} 
This representation can be labeled by a Young diagram with $s=\sum_i ia_i$ 
boxes in the 
usual way. Equivalently, we can assign to the highest weight \hw\ 
an unordered partition of $s$:
\eqn\parthw{
 a= (a_1+\cdots +a_{N-1}, a_2+\cdots + a_{N-1}, \cdots, a_{N-1},0).
}   
The weight space of this 
representation can always be written as follows 
\eqn\set{
M_{\Lambda}=\bigcup_{k_{\lambda}} m^{\Lambda}_{(k_{\lambda})} M_{k_{\lambda}},}
where the $m^{\Lambda}_{(k_{\lambda})}$ are nonnegative integers giving the 
multiplicities of the weights \wpro\ in $M_{\Lambda}$. This can be proved as 
follows (see \fh\ for more details). The irreducible 
representation associated to $\Lambda$ is given by ${\bf S}_{a}(V)$, 
where ${\bf S}_a$ is the Schur-Weyl functor. Any endomorphism of $V$ will 
extend to ${\bf S}_{a}(V)$, and its character will be given by the Schur 
polynomial $S_{a}(x_1, \cdots, x_N)$, where $x_1, \cdots, x_N$ are the
eigenvalues of $g$. The Schur polynomials can be expanded in terms of the 
symmetric  polynomials $F_{k}$, which are also labeled by unordered 
partitions of $s$, $k=(k_1, 
\cdots, k_N)$ (with $k_1 \ge \cdots \ge k_N$). 
$F_k$ is the sum of the elementary monomial $X^k
=x_{1}^{k_1}\cdots x_{N}^{k_N}$ 
and all the monomials obtained from it by permuting the variables. 
The set of $X^k$ and its permutations is then labeled by ordered partitions. 
The expansion of 
the Schur polynomials is given by:
\eqn\expschur{
S_{a} =\sum_{k}{\cal N}_{ak }F_{k},}
where the ${\cal N}_{ak}$ are called the Kostka numbers. These numbers 
are nonnegative 
integers and can be also 
computed as the number of ways one can fill the diagram $a$ with $k_1$ 
1's, $k_2$ 2's, ..., $k_r$ r's in such a way that the entries in each 
row are nondecreasing and those in each column are strictly increasing. 
Since the $x_i$, $i=1, \cdots, N$, correspond to the weights $\mu_i$ of 
the fundamental 
representation, each of the monomials in $F_k$ 
corresponds to a one-dimensional 
weight space with a weight of the form \wpro. The different monomials in $F_k$ are in one-to-one correspondence with the different 
ordered 
partitions associated to the unordered partition 
$k$. We have then proved the equality \set. 

From the  proof above follows that in the decomposition \set\ all the 
ordered partitions corresponding to the same unordered partition appear with 
the same multiplicity, and moreover that
\eqn\multi{
m^{\Lambda}_{(k_{\lambda})}= {\cal N}_{ a k},} 
where $a$ is the partition associated to $\Lambda$. We can then compute 
the multiplicities in \set\ very easily. For example, for
 $R={\rm Sym}^s(V)$, we have $\Lambda=s \lambda_1$, and 
$m^{s \lambda_1}_{(k_{\lambda})}=1$ for every ordered 
partition of $s$. For $R=\wedge^s V$ one has $\Lambda=\lambda_s$, and 
$m^{\lambda_s}_{(k_{\lambda})}=0$ 
for every $(k_{\lambda})$ except for $(k_{\lambda})=(1,1, \cdots, 1)$, 
where the multiplicity is one. For the diagram $\tableau{2 1}$, we can 
represent \set\ as:
\tableauside=1.5ex
\tableaurule=0.4pt
\eqn\three{
\tableau{2 1} = 2 (1,1,1) +(2,1) + (1,2),}
\tableauside=1.0ex
\tableaurule=0.4pt
where the vectors in the r.h.s. represent ordered partitions, and the 
coefficients are the multiplicities. 
For representations with four boxes one has:
\tableauside=1.5ex
\tableaurule=0.4pt
\eqn\four{
\eqalign{
\tableau{3 1}&= 3(1,1,1,1) + (1,3) + (3,1) + (2,2) + 
2\{ (2,1,1) + (1,2,1) + (1,1,2)\},\cr
\tableau{2 1 1}&=3(1,1,1,1) + (2,1,1) + (1,2,1) + (1,1,2),\cr
\tableau{2 2}&=2(1,1,1,1) + (2,2) + 
(2,1,1) + (1,2,1) + (1,1,2).\cr}}
\tableauside=1.0ex
\tableaurule=0.4pt

\subsec{General formula}

We are now in a position to be more explicit about the expression \vevch. 
Using the decomposition \set, we can write all the weights for the 
irreducible representation $R$ in the form \wpro. We have to find now which
vectors  of the form $ \rho + n \mu$ have a representative in ${\cal F}_l$, 
and the explicit structure of such a weight. This has been completely solved 
in Theorem 4.1 of \lm. The main output of this theorem is that the weights 
\eqn\weights{ \rho + n(k_1  \mu_{i_1} + \cdots + k_r \mu_{i_r})}
associated to a partition of cardinal $r$ 
give a representative only for certain values of 
the indices $i_1, \dots, i_r$. The procedure to get these indices, 
as well as the corresponding representative, is rather involved, but we will 
give it here for completeness. For further details, we refer the reader 
to \lm. 

The arrangement of indices ${i_{\lambda}}$, ${\lambda=1,\cdots,r},$ 
producing a weight
in ${\cal F}_l$ is contained in the set specified by the
 following conditions: 
\eqn\conds{ 
\eqalign{ 
&({\rm I})\,\ i_{\lambda} \le {k_{\lambda} n },\cr
&({\rm II})\,\ i_{\lambda}=i_{\mu}+{k_{\lambda} n },\,\,\,\ 
\mu<\lambda, \cr}
}
in such a way that, in (II), no previous index $i_{\nu}$,
$\nu<\lambda$, has the form $i_{\nu}=i_{\mu}+{k_{\nu} n }$,
$\mu<\nu$.
Given an arrangement of indices like this, with $r-k$ indices verifying
condition (I) (which will be called of type I) and $k$ indices verifying
condition (II) (which will be called of type II), a weight belonging to ${\cal
F}_l$
 is obtained if and only if:  
\eqn\condiff{
 i_{\mu}
-i_{\nu}+(k_{\nu}-k_{\mu})n \not= 0,} 
for every pair of indices $i_{\mu}$, $i_{\nu}$,  verifying (I).
The set of arrangements of indices selected in this way will be denoted by
${\cal I}_{(k_\lambda)}(n)$, and the corresponding set of weights will be 
denoted by $M_{{\cal I}_{(k_\lambda)}(n)}$.

To each arrangement of indices in ${\cal I}_{(k_\lambda)}(n)$ we will
associate a  canonical representative in ${\cal F}_l$ accompanied
by a sign. This association is carried out by the following
procedure:
 
1) For indices of type I, which will be denoted by
$i_{\lambda_1}, \cdots,i_{\lambda_{r-k}}$, one defines a total order relation
according to: 
\eqn\order{
i_{\lambda_p} \succ i_{\lambda_q}\,\ {\rm iff} \,\
i_{\lambda_p}-i_{\lambda_q}+(k_{\lambda_q}-k_{\lambda_p})n>0.
}
This relation defines a permutation $\tau$ of the set of indices
of type I under consideration with respect to their natural ordering: 
\eqn\permone{
\tau = \left(
\matrix{i_{\lambda_1}&i_{\lambda_2}&\cdots&i_{\lambda_{r-k}}\cr
i_{\tau(\lambda_1)}&i_{\tau(\lambda_2)}&\cdots&i_{\tau(\lambda_{r-k})}
\cr}\right).
}

2) For the $k$ indices of type II, $i_{
\nu_1}, \cdots , i_{ \nu_k}, \,\ i_{ \nu_1}<
\cdots < i_{ \nu_k}$, one takes the set of indices
$i_{\hat \nu_1}, \cdots , i_{\hat \nu_k}$, verifying $i_{
\nu_p}=i_{\hat \nu_p}+k_{\nu_p}n$, and defines on it the order
relation inherited from the natural ordering of the indices $i_{\nu_p}$:
\eqn\permtwo{
 i_{\hat \nu_p} \succ i_{\hat \nu_q}\,\ {\rm iff} \,\
i_{\nu_p}>i_{\nu_q}.}
This gives again a permutation $\sigma$ with respect to the natural
ordering of the set $i_{\hat \nu_p}$: 
\eqn\permthree{
\sigma=\left( \matrix {i_{\sigma^{-1}({\hat\nu_1})}&i_{\sigma^{-1}({\hat
\nu_2})}& \cdots &i_{\sigma^{-1}({\hat\nu_k})}\cr
                         i_{\hat \nu_1}& i_{\hat \nu_2}&\cdots &
i_{\hat \nu_k}\cr}\right),
}
with $i_{\sigma^{-1}({\hat\nu_1})}<i_{\sigma^{-1}
({\hat\nu_2})}<\cdots
<i_{\sigma^{-1}({\hat\nu_k})}$.

3) Define $r-k$ numbers $\xi(\lambda_p)$, $p=1,\cdots,r-k$,
associated to type I indices as follows:  $\xi(\lambda_p)$ is the
number of type II indices preceding the type I index $i_{\lambda_p}$
in the original arrangement of indices  
$i_1,\cdots, i_r$, in \weights.

The canonical representative in ${\cal F}_l$ of the weight \weights\
 is the weight: 
\eqn\repre{
\rho +  p_1 \lambda_1 + p_2 \lambda_2 + \cdots
+p_{r-k}\lambda_{r-k}+\lambda_{i_{\mu_1} +r-k-1}
+ \lambda_{i_{\mu_2}+r-k-2}+ \cdots +\lambda_{i_{\mu_{r-k}}},
}
where $p_i$, $i=1,\cdots,r-k$, are given by:
\eqn\coefrollo{
\eqalign{
p_1
=&i_{\tau(\lambda_2)}-i_{\tau(\lambda_1)}+(k_{\tau(\lambda_1)}-
k_{\tau(\lambda_2)})n - 1, \cr
p_2 =& i_{\tau(\lambda_3)} -i_{\tau(\lambda_2)} +
(k_{\tau(\lambda_2)}-k_{\tau(\lambda_3)} )n -1, \cr
&\,\,\,\,\,\,\,\,\,\,\,\,\ \vdots \cr
p_{r-k}=&{k_{\tau(\lambda_{r-k})} }n  -i_{\tau(\lambda_{r-k})},
\cr}
}
and the indices $i_{\mu_p}$ in \repre\ are the complementary ones to
the indices $\{ i_{\hat
\nu_p} \}_{p=1, \cdots, k}$, {\it i.e.}, those indices $i_{\mu_p}$
such that no index $i_{\nu}>i_{\mu_p}$ has the form $i_{\nu}=i_{\mu_p}+
k_{\nu} n$. They are ordered according to their natural ordering: $i_{\mu_1}<
\cdots <i_{\mu_{r-k}}$. 
 
Finally, the sign associated to this weight because of the Weyl reflections
needed to obtain it is: 
\eqn\signat{
\epsilon (\tau) \epsilon (\sigma) (-1)^{\sum_{p=1}^{r-k}
i_{\mu_p}-\mu_p +  \xi(\lambda_p)}.
}

This result gives then an explicit description of the set of weights 
$M(n,\Lambda)$ in \vevch: 
\eqn\finalset{
M(n,\Lambda)=\bigcup_{(k_\lambda)} m_{(k_{\lambda})}^{ \Lambda}
M_{{\cal I}_{(k_\lambda)}(n)},}
and the representatives of these weights have the form \repre.

The last ingredient in \vevch\ is the character, which has also 
been computed in \lm\ for weights with the structure of \repre. Before 
doing this, it is useful to introduce $q$-numbers and $q$-combinatorial 
numbers as follows: 
$$
[x]=t^{{x\over 2}}-t^{-{x\over 2}},\,\,\,\,\,\, (x)=t^x-1,
$$
\eqn\qtodo{
\Bigg[ {x \atop y} \Bigg]={[x]!\over [x-y]! [y]!}.}
One can then easily prove that 
\eqn\qcombn{
\eqalign{
\Bigg[ {N+p \atop  p} \Bigg]=&\lambda^{-{1\over 2}p} t^{{1  \over 2}p(p+1)} 
 {\prod_{j=-p}^{-1}
(\lambda -t^j) \over (p)!},\cr
\Bigg[{N \atop i} \Bigg]=&
\lambda^{-{i \over 2}} t^{{i \over 2}}
{\prod_{j=0}^{i-1} (\lambda-t^j) \over (i)!}.\cr}}    
The character for the weight \repre\ is given by
\eqn\character{
\eqalign{
{\rm ch}_{\Lambda}\biggl[ -{2\pi i \over k+N}\rho\biggr]
=&\prod_{k=1}^{r-1}[p_k+1] \cdots [\sum _{\lambda=k}^{r-1}
p_{\lambda}+r-1] \prod_{1 \le j<k \le r} [i_k-i_j]\cr
& \times \prod_{k=1}^r {[i_k] \over \prod_{\mu=1}^r[\sum_{\lambda=k}^r
p_{\lambda}+i_{\mu}+r-k]}\Bigg[ {N+ \sum_{\lambda=k}^r p_{\lambda}+r-k
\atop {\sum_{\lambda=k}^r p_{\lambda}+r-k}} \Bigg] \Bigg[{ N \atop
i_k} \Bigg],\cr} 
}
and it is in principle a rational function of $t^{1\over 2}$ and  
$\lambda^{1\over 2}$. 

Taking all this into account, and evaluating the phase factor in 
\vevch,  we can finally write the expression for the 
vev of a Wilson loop corresponding to an $(n,m)$ torus 
knot in an arbitrary irreducible representation $R$: 
\eqn\cocotier{ 
\eqalign{ 
&\langle W_{\Lambda}^{(n,m)}\rangle=\lambda^{ {sm (n-1) \over 2}} 
t^{{mn \over 2}\sum_i (ia_i^2 -i^2a_i)+mn\sum_{i<j}ia_i a_j-{sm \over 2}}\cr
& \sum_{(k_{\lambda})}m^{\Lambda}_{(k_{\lambda})}
t^{-{mn \over 2} \sum_{\lambda=1}^r
k_{\lambda}^2}
\sum_{{\cal I}_{(k_\lambda)}(n)}\epsilon (\tau) \epsilon  (\sigma)
(-1)^{\sum_{p=1}^{r-k} i_{\mu_p}-\mu_p+\xi(\lambda_p)}
 t^{ m \sum_{\lambda=1}^r k_{\lambda}i_{\lambda} }\cr
& \times \prod_{\tau=1}^{r-k-1}[p_{\tau}+1] \cdots [\sum
_{\lambda=\tau}^{r-k-1} p_{\lambda}+r-k-1] 
 \prod_{1 \le \sigma<\tau \le
r-k} [i_{\mu_{\tau}}-i_{\mu_{\sigma}}]\cr
& \times \prod_{\tau=1}^{r-k} {
[i_{\mu_{\tau}}] \over \prod_{j=1}^{r-k}[\sum_{\lambda=\tau}^{r-k}
p_{\lambda}+i_{\mu_j}+r-k-\tau] } \Bigg[ {N+ \sum_{\lambda=\tau}^{r-k}
p_{\lambda}+r-k-\tau \atop \sum_{\lambda=\tau}^{r-k}
p_{\lambda}+r-k-\tau} \Bigg] \Bigg[ {N \atop i_{\mu_{\tau}} } \Bigg],
\cr} }
where the highest weight associated to $R$ has been written as in \hw, 
and the multiplicities are given by Kostka numbers. This expression is 
fairly complicated. It can be evaluated using a simple computer routine, 
and in some simple cases it can be explicitly computed, as we will see in 
the next section.

\subsec{Some particular cases. Akutsu-Wadati polynomials}

In this subsection, we will study some 
particular cases of the general formula \cocotier. In particular, 
we will see that for gauge group $SU(2)$ it reduces to the 
Akutsu-Wadati polynomials of torus knots first obtained in \ilr. 
 
As a first check of \cocotier, and for completeness,
 let us consider the fundamental 
representation of $SU(N)$, with Young tableau $\tableau{1}$ and $s=1$. The 
vev in this case was obtained in the context of Chern-Simons gauge theory in
\lm.
 There 
is only one index $1\le i \le n$ involved in the procedure described 
above. The 
representatives of the weights that appear in the computation are, according to 
\repre\ and \coefrollo,
\eqn\fwrep{
\rho + (n-i)\lambda_1 + \lambda_i,}
and the character in this case is simply:
\eqn\char{
{\rm ch}_{p\lambda_1+\lambda _{i}}={[i] \over [p+i]} \Bigg [
{N+p \atop  p} \Bigg] \Bigg[{N \atop i} \Bigg ].}
Using the explicit expressions \qcombn, we finally 
obtain
\eqn\vevfund{
\langle W_{\tableau{1}}^{(n,m)} \rangle 
=t^{{1\over 2}}\lambda^{-{1\over 2}} 
{(\lambda t^{-1})^{(m-1)(n-1)\over 2}\over t^n-1}
\sum_{p+i+1=n \atop  p, i \ge 0}(-1)^i t^{-mi+{1 \over
2} p(p+1) }{\prod _{j=-p}^i (\lambda -t^j) \over (i)! (p)!},}
where we have redefined the index $i$. This is in fact the unnormalized 
HOMFLY polynomial of an $(n,m)$ torus knot. If we divide by the 
vev of the unknot, we find the expression for the HOMFLY 
polynomial first obtained in \jonesann.

An interesting check of \cocotier\ is that, for $SU(2)$ and representations 
with isospin $j/2$, one obtains in fact the Akutsu-Wadati 
polynomials for torus 
knots. The discussion is
 very similar to the reduction 
of the HOMFLY polynomial for torus links to the Jones polynomial studied 
in \lm, section 5. 

To obtain the Akutsu-Wadati polynomial for isospin $j/2$, we consider the 
representation of $SU(N)$ associated to ${\rm Sym}^j(V)$, and given
 by the Young tableau 
\tableauside=1.5ex
\tableaurule=0.4pt
$$
\tableau{7}
$$
\tableauside=1.0ex
\tableaurule=0.4pt
with $j$ boxes. We then put $N=2$.
 Notice that there are other 
representations of $SU(N)$ that reduce to the $j/2$ for $SU(2)$, but the 
above is the simpler one. From the expression for 
the character \character\ we see that it vanishes for $i_k >2$. 
This implies,  
in particular,  
that the only partitions contributing in this limit have at most 
cardinal two. From the partition $(j)$, we obtain the following 
weights in ${\cal F}_l$: 
$jn\lambda_1$ (with sign $-1$) and $(jn-2)\lambda_1 +\lambda_2$ (with sign 
$+1$). For the partitions $(l,j-l)$, with $1\le l\le j-1$, one finds 
weights of the form $p\lambda_1+ q\lambda_2$, where 
$p=n(2l-j)$, $q=n(j-l)$ if $1+n(2l-j)>0$, and $p=(j-2l)n-2$, $q=nl+1$ if 
$1+n(2l-j)<0$. In the first case, the sign is $+1$, and it is 
$-1$ in the second 
case. The character of this weight for $\lambda=t^2$ is simply
\eqn\simplechar{
{\rm ch}_{p\lambda_1 + q\lambda_2}=t^{-{p\over 2}} {t^{p+1}-1 
\over t-1}.} Taking into account that $m_{(k_{\lambda})}^{j\lambda_1}=1$ 
for all the ordered partitions, one finds, after a short calculation:
\eqn\aw{
\langle W_j^{(n,m)}\rangle =
t^{-{j\over 2}}{t^{{j\over 2}(n-1)(m-1)} \over t-1}
\sum_{l=0}^j t^{m(1+nl)(j-l)}(t^{1+nl} -t^{n(j-l)}),}
where the summands with $l=0,j$ come from the partition $(s)$, 
and the summands with $1\le l\le j-1$ come from the partition $(l,j-l)$. 
The expression in \aw\ is in fact the unnormalized Akutsu-Wadati polynomial 
for the $(n,m)$ torus knot, in the representation of isospin $j/2$ \ilr.

\newsec{Explicit results for $f_R$}

The results of the previous section will allow us to compute the quantities
on the left hand side of the master equation \master. These quantities are
products of traces of powers of the holonomy associated to a given knot.
Computations of this type are delicate from a field theory point of view
because they involve products of operators evaluated for the same loop. The
corresponding calculations are plagued with singularities which must be
regularized. One way to do this, advocated in \guada\ and 
also suggested in \ov, involves the use of
Frobenius formula. In particular, what is needed is the inverse of \frob: 
\eqn\frinv{
\Upsilon_{\vec k}(U) =\sum_R \chi_R (C(\vec k)) {\rm Tr}_R \, (U).}
All problems arising from products of operators evaluated for 
the same loop are
avoided using this formula since one ends computing vevs 
of standard Wilson loops. Actually, it is rather simple to prove that the
choice \frinv\ leads to the following general form for the functions
$f_R(t,\lambda)$:
\eqn\lot{
f_R(t,\lambda)=\langle {\rm Tr}_R (U) \rangle +
 {\rm lower \, order \, terms},}
where ``lower order terms" involves $f_{R'}(t, \lambda)$ for representations 
$R'$ with $\ell'<\ell$. Thus the set of functions
$f_R(t,\lambda)$ is equivalent to the set of vevs 
in arbitrary irreducible representations. This relation implies that the new
polynomial invariants are basically the ordinary ones plus correction terms. As
it follows from the master equation \master\ these corrections terms are linear
combinations of products of lower-order $f_R$ evaluated at different arguments.
The remarkable consequence that follows from the validity of the conjecture
\ovconj\ is that the corrected polynomials possess integer coefficients which can
be interpreted as the solutions to counting problems in the context of string
theory.

Using \frinv\ and 
the result for ${\rm Tr}_R \, (U)$ in \cocotier\ we will be able to obtain the
functions $f_R(t, \lambda)$ for torus knots after solving the master equation
\master. We will present in this section the computations up to third order, where
the order is set by $\ell$,  as we explained in section 2.

\subsec{$\ell=1$}

In this case, $\vec k =(1,0,\cdots)$ and \firstex\ just says that
\eqn\primer{
\langle {\rm Tr}_{\tableau{1}}\, U \rangle =f_{\tableau{1}}(t, \lambda).}
The left hand side of this equation is the unnormalized HOMFLY 
polynomial. To normalize it we have to divide it by the vev of the unknot:
\eqn\unknot{
\langle {\rm Tr}_{\tableau{1}}\, U \rangle_{\rm u}={\lambda^{1\over 2} 
-\lambda^{-{1\over 2}} \over t^{1\over 2} -t^{-{1\over 2}}}.}
Due to the skein relations, the normalized HOMFLY polynomial always has the
structure 
\lick: 
\eqn\homstr{
{\langle {\rm Tr}_{\tableau{1}}\, U \rangle \over 
\langle {\rm Tr}_{\tableau{1}}\, U \rangle_{\rm u}}= 
\sum_{s} p_s(\lambda)t^s. }
In this equation, the $s$ take integer values, 
and $p_s(\lambda)=\sum_j a_{s,j}\lambda^j$ are Laurent polynomials in 
$\lambda$. The $a_{s,j}$ are integer 
numbers. Therefore, 
$f_{\tableau{1}}(t,\lambda)$ has indeed the structure predicted in \ovconj. 
The integers $N_{\tableau{1},Q,s}$ are given by:
\eqn\integers{
N_{\tableau{1},j+1/2,s}=a_{s,j}-a_{s,j+1}.}
We then see that, for the fundamental 
representation, the integers introduced in \ov\ are simple linear 
combinations of the coefficients in the normalized HOMFLY polynomial. Notice that 
\integers\ is valid for any knot, since we have only used the 
general structure of the HOMFLY polynomial. As an example, 
let us consider the right-handed trefoil, which is the $(2,3)$ torus knot. One
obtains that
\eqn\trefund{
f_{\tableau{1}} (t,\lambda)={1 \over t^{1\over 2}
 -t^{-{1\over 2}} }(-2  \lambda^{1\over2} + 3  \lambda^{3\over2} 
- \lambda^{5\over2}) + (t^{1\over 2} - t^{-{1\over 2}})(-\lambda^{1\over2} +  
\lambda^{3\over2}),}
and from here one can easily extract the values of $N_{\tableau{1},Q,s}$.

\subsec{$\ell=2$}   
In this case, there are two possible vectors corresponding to conjugacy 
classes: $(2,0,\cdots)$, and $(0,1,0,\cdots)$. From \master, 
\firstex\ and \secex\ we find two equations:
\eqn\ordertwo{
\eqalign{
\langle ({\rm Tr}\,U)^2\rangle-\langle {\rm Tr}\,U\rangle ^2 =
&f_{\tableau{2}}(t, \lambda) +f_{\tableau{1 1}}(t, \lambda),\cr 
\langle {\rm Tr}\,U^2\rangle=&f_{\tableau{2}}(t, \lambda) 
-f_{\tableau{1 1}}(t, \lambda)+f_{\tableau{1}}(t^2, \lambda^2).\cr}}
A432s argued above, Frobenius formula \frinv\ allows to express the new quantities
appearing on the left of this equation in terms vevs of Wilson loops. For the case
under consideration it leads to:
\eqn\frodos{
\eqalign{
\langle ({\rm Tr}\,U)^2\rangle =& \langle {\rm Tr}_{\tableau{2}}\,U\rangle +
\langle {\rm Tr}_{\tableau{1 1}}\,U\rangle,\cr
\langle {\rm Tr}\,U^2\rangle =&\langle {\rm Tr}_{\tableau{2}}\,U\rangle -
\langle {\rm Tr}_{\tableau{1 1}}\,U\rangle.\cr}}
From these relations and eq. \ordertwo\ we obtain, after taking into account
\primer:
\eqn\ftwos{
\eqalign{
f_{\tableau{2}}(t, \lambda)=&\langle {\rm Tr}_{\tableau{2}}\,U\rangle
-{1\over 2}\bigl( f_{\tableau{1}}(t,\lambda)^2+ 
f_{\tableau{1}}(t^2,\lambda^2) 
\bigr),\cr
f_{\tableau{1 1}}(t, \lambda)=&\langle {\rm Tr}_{\tableau{1 1}}\,U\rangle
-{1\over 2}\bigl( f_{\tableau{1}}(t,\lambda)^2-
 f_{\tableau{1}}(t^2,\lambda^2) 
\bigr).\cr}}

We will now present explicit formulae for these functions in the 
simplest nontrivial case, namely the right-handed trefoil knot. The vev in the
symmetric representation is given by:
\eqn\vevsymtre{
\langle {\rm Tr}_{\tableau{2}} \, U \rangle=
 {(\lambda-1)(\lambda t-1)  \over
 \lambda ( t^{{1\over 2}} -
 t^{-{1\over 2}}) ^2 \,( 1 + t)}
\Bigl((\lambda t^{-1})^2
( 1 - {\lambda}t^2 +t^3-\lambda t^3+ t^4 -\lambda t^5 
+ \lambda^2 t^5 +t^6 -\lambda t^6) \Bigr).}
In this equation, we have explicitly factored out 
the vev for the unknot in the symmetric 
representation. The polynomial 
multiplying the fraction in the right hand side is then 
the normalized polynomial invariant in the symmetric 
representation. One can see that this expression agrees with the 
result presented in \rama. It can also be easily checked that, when we 
substitute $\lambda \rightarrow t^2$, we obtain the 
Akutsu-Wadati polynomial for the right-handed trefoil in the $j=2$
representation,  as it should be. 

For the antisymmetric representation we find: 
\eqn\vevantytre{
\langle {\rm Tr}_{\tableau{1 1}} \, U \rangle=
{(\lambda-1)(\lambda-t) \over \lambda  (  t^{{1\over 2}} -
 t^{-{1\over 2}}) ^2\,\,( 1 + t )  } 
\Bigl( (\lambda t^{-2})^2 (1 -{\lambda} - \lambda t + {{{\lambda}}^2} t
+ t^2 +t^3 - {\lambda}t^3 -{{\lambda}}\,t^4 +t^6 ) \Bigr). } 
Notice that, when $N=2$ ({\it i.e.}, when $\lambda=t^2$), $\tableau{1 1}$ 
becomes the trivial representation and indeed \vevantytre\ becomes $1$. 

Using \ftwos\ one finds:
\eqn\efestre{\eqalign{
f_{\tableau{2}}(t,\lambda)
&={ t^{-{1\over 2}}{\lambda}( {\lambda}-1) ^2 \,\,( 1 + {t^2}) \,
     ( t + {{{\lambda}}^2}\,t - {\lambda}\,( 1 + {t^2} ))
\over t^{{1\over 2}} - t^{-{1\over 2}} }  
\cr
f_{\tableau{1 1}}(t,\lambda)& = - {1 \over t^3}f_{\tableau{2}}(t,\lambda).} 
}
The structure of these functions is in perfect agreement with
 \ovconj. This computation makes clear that the prediction \ovconj\ 
is far from being trivial from the Chern-Simons side. The vevs \vevsymtre\ 
and \vevantytre\ have complicated denominators that have to cancel out 
except for a single factor of $t^{{1\over 2}} - t^{-{1\over 2}}$ when one 
substracts the lower order contributions as in \ftwos. Also notice that 
the coefficients of the functions in \efestre\ are in fact integers, and 
again this is not obvious from \ftwos\ (which involves dividing by $2$). 
These features become more and more remarkable as we increase 
the number of boxes of the representations, as we will see.

\subsec{$\ell=3$}
At this order there are three vectors that contribute, $\vec k=(3,0,\cdots)$, 
$\vec k=(1,1,0,\cdots)$, and $\vec k=(0,0,1,0,\cdots)$. From \firstex\ we 
obtain: 
\eqn\orderthree{
\langle ({\rm Tr}\,U)^3\rangle-3\langle {\rm Tr}\,U\rangle \langle 
({\rm Tr}\, U)^2 \rangle +2 \langle {\rm Tr}\, U \rangle^3=
 f_{\tableau{3}}(t, \lambda) +f_{\tableau{1 1 1}}(t, \lambda) 
+ 2f_{\tableau{2 1}}(t, \lambda),}
while from \secex\ one has:
\eqn\porderthree{
\langle {\rm Tr}\, U^3 \rangle = f_{\tableau{3}}(t, \lambda)
 +f_{\tableau{1 1 1}}(t, \lambda) 
-f_{\tableau{2 1}}(t, \lambda) + f_{\tableau{1}}(t^3, \lambda^3).}
Finally, the vector $(1,1,0,\cdots)$ gives us: 
\eqn\finalthree{
\langle {\rm Tr}\,U \, {\rm Tr}\, U^2\rangle-\langle {\rm Tr}\,U\rangle 
\langle {\rm Tr}\,U^2\rangle = 
f_{\tableau{3}}(t, \lambda)-f_{\tableau{1 1 1}}(t, \lambda).}
Using again Frobenius formula, we find:
\eqn\orderexp{
\eqalign{
f_{\tableau{3}}(t, \lambda)&=
\langle {\rm Tr}_{\tableau{3}}\,U \rangle 
-f_{\tableau{1}}(t,\lambda) f_{\tableau{2}}(t,\lambda) 
-{1\over 6}f_{\tableau{1}}(t,\lambda)^3 \cr
&- {1\over 2}f_{\tableau{1}}(t,\lambda) f_{\tableau{1}}(t^2,\lambda^2)
-{1\over 3} f_{\tableau{1}}(t^3,\lambda^3),\cr
f_{\tableau{2 1}}(t,\lambda)&=
\langle {\rm Tr}_{\tableau{2 1}}\,U \rangle 
-f_{\tableau{1}}(t,\lambda) (f_{\tableau{2}}(t,\lambda)+
f_{\tableau{1 1}}(t,\lambda))
-{1\over 3}f_{\tableau{1}}(t,\lambda)^3
+{1\over 3}f_{\tableau{1}}(t^3,\lambda^3),\cr
f_{\tableau{1 1 1}}(t,\lambda)&=
\langle {\rm Tr}_{\tableau{1 1 1}}\,U \rangle
-f_{\tableau{1}}(t,\lambda) f_{\tableau{1 1}}(t,\lambda) 
-{1\over 6} f_{\tableau{1}}(t,\lambda)^3 
+{1\over 2} f_{\tableau{1}}(t,\lambda) f_{\tableau{1}}(t^2,\lambda^2) 
-{1\over 3} f_{\tableau{1}}(t^3,\lambda^3).\cr}}

Let's now present some results for the right-handed trefoil knot. For the
representation 
$\tableau{3}$, we find:
\eqn\treftre{
\eqalign{
\langle {\rm Tr}_{\tableau{3}}\,U \rangle =
&{ (\lambda -1)(\lambda t-1) (\lambda t^2-1)
\over \lambda^{3/2}(t^{1\over 2} -t^{-{1\over 2}})^3
 \left( 1 + t \right) \,\left( 1 + t + {t^2} \right) 
}\Bigl( (\lambda t^{-1})^3 
(1 - {\lambda}\,{t^3} + {t^4} - {\lambda}\,{t^4} + {t^5} - 
{\lambda}\,{t^5}\cr
& + {t^6} - {\lambda}\,{t^7} + 
  {{{\lambda}}^2}\,{t^7} + {t^8} - 2\,{\lambda}\,{t^8} 
 + 
{{{\lambda}}^2}\,{t^8} + {t^9} - 
  2\,{\lambda}\,{t^9} + {{{\lambda}}^2}\,{t^9} + {t^{10}} -
 {\lambda}\,{t^{10}}\cr  
&  - {\lambda}\,{t^{11}} + 
  {{{\lambda}}^2}\,{t^{11}}+ {t^{12}} - {\lambda}\,{t^{12}} + {{{\lambda}}^2}\,{t^{12}} - 
  {{{\lambda}}^3}\,{t^{12}} - {\lambda}\,{t^{13}} + 
{{{\lambda}}^2}\,{t^{13}})\Bigr),\cr}
} 
which also agrees with the computation in \rama. 
For the representation $\tableau{2 1}$, we find: 
\eqn\mixta{
\eqalign{
\langle {\rm Tr}_{\tableau{2 1}}\,U \rangle =
 &{ (\lambda -1)(\lambda -t) (\lambda t -1)
\over \lambda^{3/2}(t^{1\over 2} -t^{-{1\over 2}})^3
 \,\left( 1 + t + {t^2} \right) 
}\Bigl( \lambda^3 t^{-5} (
1 - {\lambda} + 2\,{t^2} - 2\,{\lambda}\,{t^2} + {{{\lambda}}^2}\,{t^2}
\cr & - {t^3} + 
  {{{\lambda}}^2}\,{t^3} + 2\,{t^4} - 3\,{\lambda}\,{t^4} + 
{{{\lambda}}^2}\,{t^4} - 
  {{{\lambda}}^3}\,{t^5} + 2\,{t^6} - 3\,{\lambda}\,{t^6} + 
{{{\lambda}}^2}\,{t^6} - {t^7} \cr &+  
  {{{\lambda}}^2}\,{t^7} + 2\,{t^8} - 2\,{\lambda}\,{t^8} + {{{\lambda}}^2}\,{t^8} + {t^{10}} - 
  {\lambda}\,{t^{10}} ) \Bigr).\cr}}
Notice that, when $\lambda \rightarrow t^2$, the normalized polynomial 
(which is the polynomial inside the parentheses in \mixta) goes to the 
Jones polynomial of the right-handed trefoil, since the representation
$\tableau{2 1}$  reduces to the fundamental representation $j=1$ when $N=2$. 

Finally, for the representation $\wedge^3 V$, with Young diagram 
$\tableau{1 1 1}$, one has:
\eqn\mixta{
\eqalign{
\langle {\rm Tr}_{\tableau{1 1 1}}\,U \rangle =
 &{ (\lambda -1)(\lambda -t) (\lambda -t^2)
\over \lambda^{3/2}(t^{1\over 2} -t^{-{1\over 2}})^3
(1+t) \,\left( 1 + t + {t^2} \right) 
}\Bigl( \lambda^3 t^{-10} (- 
{\lambda} + {{{\lambda}}^2} + t - {\lambda}\,t +
 {{{\lambda}}^2}\,t \cr & - {{{\lambda}}^3}\,t - 
  {\lambda}\,{t^2} + {{{\lambda}}^2}\,{t^2} + {t^3} -
 {\lambda}\,{t^3} + {t^4} - 2\,{\lambda}\,{t^4} + 
  {{{\lambda}}^2}\,{t^4} + {t^5} - 2\,{\lambda}\,{t^5}\cr &  +
 {{{\lambda}}^2}\,{t^5} - {\lambda}\,{t^6} + 
  {{{\lambda}}^2}\,{t^6} + {t^7} + {t^8} - {\lambda}\,{t^8} + 
{t^9} - {\lambda}\,{t^9} -
  {\lambda}\,{t^{10}} + {t^{13}}) \Bigr).\cr }}
Using these vevs, one finds:
\eqn\fthreetre{
\eqalign{
f_{\tableau{3}}(\lambda,t)=&-{\lambda^{3\over 2}t^{-1}(\lambda -1)^2
\over t^{1\over 2} -t^{-{1\over 2}}}
\Bigl( {t^3}\,\left( 1 + t + {t^3} \right)  + {{\lambda}^4}\,{t^3}\,\left( 1 + t + {t^2} + {t^3} + {t^4} + {t^6} \right) 
\cr & - 
  \lambda\,t\,\left( 1 + 3\,t + 3\,{t^2} + 4\,{t^3} + 5\,{t^4} +
 2\,{t^5} + 2\,{t^6} + {t^7} \right) \cr & - 
  {{\lambda}^3}\,t\,\left( 1 + 3\,t + 3\,{t^2} + 5\,{t^3} +
 5\,{t^4} + 4\,{t^5} + 2\,{t^6} + 3\,{t^7} + {t^9} \right)
     \cr & + {{\lambda}^2}\,\left( 1 + 2\,t + 3\,{t^2} +
 7\,{t^3} + 7\,{t^4} + 6\,{t^5} + 6\,{t^6} + 4\,{t^7} + {t^8} + 
     2\,{t^9} \right)\Bigr), \cr
f_{\tableau{2 1}}(t,\lambda)=
&{\lambda^{3\over 2}t^{-4}(\lambda -1)^2 (1+t+t^2)
\over t^{1\over 2} -t^{-{1\over 2}}}
\Bigl({t^3} + {\lambda^4}\,( {t^2} + {t^4} )  - 
  \lambda\,t\,( 1 + 2\,t + {t^2} + 2\,{t^3} + {t^4})\cr
 &  - {\lambda^3}\,t\,( 2 + t + 3\,{t^2} + {t^3} + 2\,{t^4} )  + 
  {\lambda ^2}\,( 1 + t + 3\,{t^2} + 3\,{t^3} + 3\,{t^4} +
 {t^5} + {t^6} )\Bigr),\cr
f_{\tableau{1 1 1}}(t,\lambda)=&-{(\lambda^{1\over 2}t^{-3})^3 
(\lambda -1)^2
\over t^{1\over 2} -t^{-{1\over 2}}}\Bigl( 
{t^4} + {t^6} + {t^7} + 
{{\lambda}^4}\,\left( t + {t^3} + {t^4} + {t^5} + {t^6} + {t^7} \right)
\cr &  - 
  \lambda\,{t^2}\,\left( 1 + 2\,t + 2\,{t^2} + 5\,{t^3} + 4\,{t^4} +
 3\,{t^5} + 3\,{t^6} + {t^7} \right)\cr &  + 
  {{\lambda}^2}\,t\,\left( 2 + t + 4\,{t^2} + 6\,{t^3} + 6\,{t^4} +
 7\,{t^5} + 7\,{t^6} + 3\,{t^7} + 2\,{t^8} + 
     {t^9} \right) \cr & - 
{{\lambda}^3}\,\left( 1 + 3\,{t^2} + 2\,{t^3} + 4\,{t^4} + 5\,{t^5} + 5\,{t^6} + 3\,{t^7} + 
     3\,{t^8} + {t^9} \right)\Bigr).
\cr}}
Again, this is in perfect agreement with \ovconj. In the appendix we list 
the $f_R(t, \lambda)$ for the right-handed trefoil knot, for representations with 
four boxes.

\subsec{Structure of $f_R$}

The functions $f_R(t,\lambda)$ that we have listed in this section, 
as well as many other examples that we have explicitly computed, 
have the structure predicted by \ovconj. In all cases they can be 
written as
\eqn\strudestru{
f_R(t,\lambda)={\lambda^{1\over 2} 
-\lambda^{-{1\over 2}} \over t^{1\over 2} -t^{-{1\over 2}}}
P_R(t^{1\over 2},\lambda^{1\over 2}), } 
where $ P_R(t^{1\over 2},\lambda^{1\over 2})$ is a Laurent polynomial
in $t^{1\over 2}$, $\lambda^{1\over 2}$ 
with integer coefficients. Notice that we have factored out the 
vev of the unknot in the fundamental representation. 
The above structure is far from being obvious from its 
definition, or from the explicit expressions given above: to get 
$f_R(t,\lambda)$ we have to add up functions with rather complicated 
denominators, however the result has the simpler 
structure given in \strudestru. Also, to obtain $f_R(t,\lambda)$ we have 
to divide by $\ell!$, however the coefficients of the resulting 
polynomial in \strudestru\ have integer coefficients. 
Notice that \strudestru\ has an extra piece of information when compared 
to \ovconj: namely, that one can extract a common factor $\lambda^{1\over 2} 
-\lambda^{-{1\over 2}}$ from the functions $f_R(t, \lambda)$. It would be 
interesting to see if this is a general fact, and if it can be 
also predicted from the string side. 

It also follows from our computations that, for a given irreducible representation 
$R$, the integers $N_{R,Q,s}$ are only different from zero for a finite 
number of values of $Q$ and $s$. However, the functions $f_R$ become more 
and more complicated as we increase the number of boxes, even for the 
trefoil knot (which is the simplest nontrivial knot). This seems 
to indicate that,  
given an irreducible representation $R$, there are always values of $Q$ and $s$ for
which 
$N_{R,Q,s} \not= 0$. Therefore, for every nontrivial knot there seems 
to be an 
infinite number of nonzero integers $N_{R,Q,s}$.

\subsec{The functions $f_R$ for the unknot}

In \ov\ it was explicitly shown that, for the unknot, 
the functions  $f_R(t,\lambda)$ vanish for all $R$ but the 
fundamental representation. This
property can be easily checked using the fact that for the unknot,
\eqn\devi{
\langle {\rm Tr}_{R}\,U \rangle_{\rm u} = {\rm dim}_t R,}
where ${\rm dim}_t R$ is the quantum dimension of the representation $R$. Recall
that this dimension is easily computed for $SU(N)$ using the hook rule to
calculate the ordinary dimension of $R$. This rule assigns a quotient to ${\rm
dim}\, R$ obtained in the following way: for the numerator, products  of $N\pm
i$,
$i=0,1,2,\dots$, each coming from a box located on the parallel to the diagonal
placed $\pm i$ times away from the diagonal, taking the plus sign for the upper
part and the minus sign for the lower part; a denominator provided by the hook
lengths. For example, for the Young tableau:
\tableauside=1.5ex
\tableaurule=0.4pt
\eqn\ejem{R = \tableau{2 1},}
\tableauside=1.0ex
\tableaurule=0.4pt
the dimension is,
\eqn\ladi{{\rm dim} R = {N(N+1)(N-1) \over 3\cdot 1 \cdot 1}.}
The corresponding quantum dimension is obtained after replacing each of the
integers $n$ appearing in the quotient by its corresponding quantum number,
\eqn\ququ{\{x\}={t^{x\over 2}-t^{-{x\over 2}} \over t^{1\over 2}-t^{-{1\over
2}}},} so that,
\eqn\ladiqu{{\rm dim}_t R = {\{N\}\{N+1\}\{N-1\} \over \{3\} \{1\} \{1\}}.}

Using \primer\ we obtain for the unknot,
\eqn\lisa{f_{\tableau{1}}(t, \lambda)_{\rm u} = 
\langle {\rm Tr}_{\tableau{1}}\, U \rangle_{\rm u} =
{\rm dim}_t {\tableauside=1.5ex\tableau{1}\tableauside=1.0ex} = \{N\},}
which is consistent with \unknot. This relation is also consistent with the
results in \ov. Let us now test the rest of the expressions for the functions
$f_R$ which we have obtained. Taking \ftwos\ one finds that, indeed,
\eqn\ftwoscheck{
\eqalign{
f_{\tableau{2}}(t, \lambda)_{\rm u}={\rm dim}_t
{\tableauside=1.5ex\tableau{2}\tableauside=1.0ex} -{1 \over
2}\Bigl(\{N\}^2+\{N\}\Big|_{t\rightarrow t^2}\Bigr)=0,\cr f_{\tableau{1 1}}(t,
\lambda)_{\rm u}={\rm dim}_t {\tableauside=1.5ex\tableau{1
1}\tableauside=1.0ex} -{1 \over 2}\Bigl(\{N\}^2-\{N\}\Big|_{t\rightarrow
t^2}\Bigr)=0.\cr}} Similarly, using these results and \orderexp\ one confirms
that for the functions of third order:
\eqn\fthreescheck{
f_{\tableau{3}}(t, \lambda)_{\rm u}=f_{\tableau{2 1}}(t, \lambda)_{\rm u}=
f_{\tableau{1 1 1}}(t, \lambda)_{\rm u}=0.}
Eqs.  \ftwoscheck\  and \fthreescheck\   constitute an important check of our
calculations.

\subsec{Perturbative expansions and Vassiliev invariants}

Using the same arguments as in \lpp\ to prove that the coefficients of the
perturbative series expansion are Vassiliev invariants \vass\ one can easily show
that the vevs \vevs\ lead to a perturbative series expansion whose coefficients
are also Vassiliev invariants. This implies that the functions $f_R(t, \lambda)$
share the same properties. In other words, if one considers the power series
expansion,
\eqn\pws{
f_R({\rm e}^x,{\rm e}^{N x})=\sum_{i=0}^{\infty} \alpha_i x^i,}
the coefficients $\alpha_i$, $i=0,1,2,\dots$, are Vassiliev invariants of order
$i$. The explicit form of the Vassiliev invariants for torus knots $(n,m)$ are
known \al\simon\ up to order six. They turn out to be polynomials in $n$ and $m$.
At lowest orders, the form of these invariants imply the following structure for
the polynomials $P_R$ in \strudestru:
\eqn\tania{P_R({\rm e}^x,{\rm e}^{N x})=
g_0+g_2(R) \beta_{2,1} x^2 + g_3(R) \beta_{3,1} x^3 + {\cal O}(x^4),}
where,
\eqn\lola{
\eqalign{
\beta_{2,1} =& {1\over 24} (n^2-1)(m^2-1), \cr
\beta_{3,1}=& {1 \over 144} n m (n^2-1) (m^2-1), \cr}}
and $g_2(R)$ and $g_3(R)$ are constants (independent of $n$ and $m$) which
depend on the representation $R$. After computing $P_R$ for a variety of torus
knots we find,
\eqn\pertu{
\eqalign{
P_{\tableau{2}}(x)&= 2N(N^2-1) \beta_{3,1} x^3 + {\cal O}(x^4),\cr 
P_{\tableau{1 1}}(x)&= -2N(N^2-1) \beta_{3,1} x^3 + {\cal O}(x^4),\cr
P_{\tableau{3}}(x)&=6N(N^2-1) \beta_{3,1} x^3 + {\cal O}(x^4),\cr
P_{\tableau{2 1}}(x)&=-6N(N^2-1) \beta_{3,1} x^3 + {\cal O}(x^4),\cr
P_{\tableau{1 1 1}}(x)&=6N(N^2-1) \beta_{3,1} x^3 + {\cal O}(x^4),\cr
}}
in full agreement with \tania.

These results constitute a test of the fact that the coefficients of the
perturbative series expansion associated to the polynomials $P_R$ must be
Vassiliev invariants. But the test indicates the existence of more structure. As
argued above, the functions $f_R$ have a very simple structure, many
cancellations occur in such a way that these functions have a simple denominator
and the vev for the unknot factorizes. The results \pertu\ indicate that they
might satisfy more striking properties. Though the fact that $g_0=0$ is a simple
consequence of 
$\ftwoscheck$ and $\fthreescheck$, there is no reason to expect that $g_2(R)=0$
for the representations under consideration. This property implies that the
second derivative respect to $t$ of $P_R$ (after replacing $\lambda \rightarrow
t^N$) vanishes at $t=1$. This might be a first indication of the existence of
some important properties shared by the polynomials $P_R$. Further work is
needed to study their general features and applications. In particular, it 
would be very interesting to understand in more detail the relation between 
the coefficients $N_{R,Q,s}$ of \ovconj\ and Vassiliev invariants.

\newsec{A conjecture for the connected vevs}

In the previous sections we have shown how to extract string amplitudes 
from Chern-Simons vevs, and that the amplitudes 
computed in that way have in fact the structure predicted in \ovconj\ 
by using the target space interpretation. The worldsheet interpretation of 
the amplitudes is in principle more complicated, since it involves 
open string instantons. However, the structure of the free energy of 
topological string theory dictated by worldsheet perturbative considerations 
gives a remarkable set of constraints on the integers $N_{R,Q,s}$ or, 
equivalently, on the connected vevs of Chern-Simons gauge theory.
 As explained in \ov, the arguments in \witop\ imply 
that the free energy $F(V)=-\log \langle 
Z(U,V)\rangle$ is given by the expression
\eqn\mini{
F(V)=\sum_{g=0}^\infty \sum_{h=1}^\infty \sum_{n_1, \cdots, n_h} 
x^{2g-2 +h}F_{g;n_1, \cdots, n_h}(\lambda)\, {\rm Tr}\,V^{n_1} 
\cdots {\rm Tr}\, V^{n_h},}
where $x$ is $2\pi i /(k+N)$, and $g$ and $h$ denote the genus 
and number of boundaries 
of the string worldsheet. Let us now compare this expression with the 
equations \conj\ and \ovconj, based on the target space interpretation. 
To do this we assume that there is some analytical continuation which 
turns the series \mini\ into a series involving only positive integers $n_i$. 
If we then choose the basis \basis\ for the class functions, we find:
\eqn\frek{
F(V)= \sum_{\vec k}{|\vec k|! \over \prod k_j!} 
\sum_{g=0}^{\infty} F_{g, \vec k}(\lambda) x^{2g-2+ |\vec k|} 
\Upsilon_{\vec k} (V).} 
Comparing to \lojari\ we immediately obtain:
\eqn\weakpre{
G_{\vec k}^{(c)}(U)=-{|\vec k|! \prod j^{k_j}}
\sum_{g=0}^\infty F_{g,\vec k}(\lambda)\,x^{2g-2+|\vec k|}.}
This makes a highly nontrivial prediction about the structure of the 
connected vevs: if we put $t={\rm e }^x$, keeping $\lambda$ as 
an independent variable, then the expansion in $x$ should start 
with a power greater or equal than $|\vec k|-2$. Moreover, the 
expansion should contain powers of the same parity ({\it i.e.} the powers 
should be all even or all odd, depending on the parity of $|\vec k|$). 
This implies that the functions $G_{\vec k}^{(c)}(U)$ are even (odd) under 
$t\leftrightarrow t^{-1}$ when $|\vec k|$ is even (odd).  

Let us now analyze this prediction. For $\ell=1$ 
(and therefore $|\vec k|=1$), the left hand side of \weakpre\ 
is the unnormalized 
HOMFLY polynomial. The fact that the expansion in $x$ starts with $x^{-1}$ 
is a consequence of \homstr. Since the normalized HOMFLY polynomial 
is even under the exchange of $t$ and $t^{-1}$ \lick, 
$G_{(1,0, \cdots)}^{(c)}(U)$ is odd, in agreement with the 
prediction. These two facts were already noted in \gvtwo\ in this context 
(indeed, the equation \weakpre\ generalizes equation (2.3) of 
\gvtwo\ to more complicated vevs). For $\ell>1$, 
the prediction \weakpre\ is far from being 
obvious: the 
vevs of Wilson loops in the representation $R$ start typically with the 
power $x^{-\ell}$ when we do not expand $\lambda$, and they do not have 
any a priori symmetry under $t \leftrightarrow t^{-1}$. However, we 
have found that the prediction \weakpre\ is in fact true in  
all the cases that we have checked. For example, in the case 
of the right-handed trefoil knot, and for the connected vevs at order four, 
we have obtained:
\eqn\vevcua{
\eqalign{
G_{(0,0,0,1,0,\cdots)}^{(c)}(U)=&{1 \over 4}\lambda^2 
(\lambda-1)(-134 + 1498 \lambda -6278 \lambda^2 
+ 13146 \lambda^3 -15129 \lambda^4\cr 
& + 9735 \lambda^5 -3289 \lambda^6 
+ 455 \lambda^7) \, {1 \over x} + {\cal O}(x),\cr 
G_{(0,2,0,\cdots)}^{(c)}(U)=&12\lambda^2(\lambda-1)^4 
(9-72\lambda + 198 \lambda^2 -176 \lambda^3 + 49 \lambda^4) + 
{\cal O}(x^2),\cr
G_{(1,0,1,0,\cdots)}^{(c)}(U)=&9\lambda^2 (\lambda-1)^4 
(10-92\lambda + 233 \lambda^2 -200 \lambda^3 + 55 \lambda^4) + 
{\cal O} (x^2),\cr 
G_{(2,1,0,\cdots)}^{(c)}(U)=
 &72(\lambda-1)^5 \lambda^2(-3+ 27\lambda - 58\lambda^2 + 
28\lambda^3)\, x +{\cal O}(x^3),\cr
G_{(4,0,0,\cdots)}^{(c)}(U)= &432 (\lambda-1)^6\lambda^2
(1-9\lambda + 16\lambda^2)\, x^2 + {\cal O}(x^4).\cr}}
In addition, one finds that the expansion only contains powers of 
$x$ of the same parity, in agreement with \weakpre. We think that 
this result gives another important check of the open string 
interpretation of Chern-Simons gauge theory. 

The prediction \weakpre\ can be stated in terms of the integer invariants 
$N_{R,Q,s}$ by using our master equation \master. Notice that, from 
\master, the most we can say about the expansion of the connected vevs 
is that they start with $x^{-1}$. However, more is true, as we have 
just seen. This means that there must be some constraints 
on the integer invariants $N_{R,Q,s}$. Let us obtain these constraints. 
 Using the definition of 
the Bernoulli polynomials,
\eqn\bern{
{{\rm e}^{xt}\over {\rm e}^t -1}=\sum_{m=0}^\infty B_m(x){t^{m-1} \over m!},}
we find the following equation for $F_{g, \vec k}$:
\eqn\fgk{
F_{g,\vec k}(\lambda)=-{1 \over |\vec k|! \prod j^{k_j}} 
\sum_{n |\vec k} n^{2g+ 2|\vec k|-3} \sum_{R,Q,s} 
\chi_R( C(\vec k_{1/n})) N_{R,Q,s}  
{B_{2g-1+ |\vec k|}(s+1/2) \over (2g-1+ |\vec k|)!}\lambda^{nQ}.}
This expression can be interpreted as a multicovering formula 
for open string instantons, in the spirit of \gvm. 
Notice that the sum over representations 
in this equation is finite, as in \master. 
The structure of the expansion in \frek\ also 
implies the following sum rules. Fix a vector $\vec k$ and a half-integer 
$j$. 
Then, one has: 
\eqn\sumrules{ 
\sum_{n |\vec k} n^{|\vec k|+m-2} \sum_{R,s} 
\chi_R( C(\vec k_{1/n})) N_{R,j/n,s} 
B_{m}(s+1/2) =0, } 
when $m \equiv |\vec k|$ mod $2$, and also when $m=0,1, \cdots, |\vec k|-2$ 
(for $|\vec k| \ge 2$). $N_{R,j/n,s}$ is taken to be 
zero if $j/n$ is not a half-integer. Notice that the sum in 
\sumrules\ involves only a finite 
number of terms. The sum rules \sumrules\ encode the properties about the 
perturbative expansion of the connected vevs that we discussed above, 
in terms of the integers $N_{R,Q,s}$.   

\newsec{Conclusions and open problems}

In this paper we have presented strong evidence for the existence of new
polynomial invariants, $f_R$, whose integer coefficients $N_{R,Q,s}$ 
can be regarded as the
solutions of certain enumerative problem in the context of string theory. 
These
polynomials are labeled by irreducible representations of $SU(N)$, and for the
fundamental representation they correspond to the unnormalized HOMFLY polynomials.
For other irreducible representations they have the form of the corresponding
unnormalized ordinary polynomial invariants, plus a series of correction terms
which involve representations whose associated Young tableaux have a lower number
of boxes. Their existence would answer a basic question in knot theory which has
remained open for many years: polynomial invariants, appropriately corrected, can
indeed be regarded as generating functions.

The evidence for the existence of the new polynomials is a consequence of the 
precision test of the  correspondence between Chern-Simons gauge theory and
topological strings  carried out in this paper. We have proved that one can in
fact extract the string amplitudes  from the Chern-Simons vevs following a
recursive procedure. This  makes possible to compute the integer invariants
$N_{R,Q,s}$ starting  from the Chern-Simons side. Using explicit results for
torus knots,  we have been able to give remarkable evidence for the predictions
of \ov,  and we have also exploited the interplay between worldsheet and target 
results to give further checks of the string theory interpretation of 
Chern-Simons gauge theory. 

There are clearly two different avenues for future research. 
On the Chern-Simons side, it would be extremely interesting to extend these 
results to more general knots, $n$-component links, and/or other gauge groups. It
would be also very interesting to explore in more detail  the relations
between the integers $N_{R,Q,s}$ and the two other sets  of integer invariants of
knots: the coefficients of the normalized  polynomials, and the Vassiliev
invariants.  On the string side, the duality with Chern-Simons gauge theory 
opens the possibility of extracting information about 
open string instantons in the resolved conifold geometry. The procedure 
we have developed in this paper gives a very concrete strategy to 
compute the string amplitudes and obtain the relevant spectrum of BPS states 
associated to D2 branes ending on D4 branes. As a preliminary step, 
one should make more precise the geometry of the Lagrangian submanifold 
in the resolved geometry. We hope to report on these 
and other related issues in the near future.

\vfill    
\eject

\bigskip
\centerline{\bf Acknowledgments}
\bigskip

We would like to thank M. Bershadsky, M. Douglas and A.V. Ramallo for useful 
conversations. We are specially indebted to H. Ooguri and C. Vafa for 
discussions and correspondence, and for a critical reading 
of the manuscript. M.M. would like to thank the Departamento de
F\'\i sica de Part\'\i culas at the Universidade de Santiago de
Compostela, where part of this work was done, for their
hospitality. The work by J.M.F.L. was supported in part by DGICYT
under grant PB96-0960. The work of M.M. is
supported by DOE grant DE-FG02-96ER40959.

\vfill    
\eject

\appendix{A}{The functions $f_R(t, \lambda)$ for $\ell=4$}
In this appendix, we list the functions $f_R (t, \lambda)$ 
for representations with four boxes, in the case of 
the right-handed trefoil knot. The results are: 
\eqn\fsymfour{
\eqalign{
 f_{\tableau{4}}(t, \lambda)=&
{t^{-3/2}\over t^{1\over 2}- t^{-{1\over 2}}}
 \left( -1 +\lambda \right) ^2\, {\lambda^2}\,
      \left( \lambda - t \right) \,
      \left( -1 + \lambda\,t \right)  \cr 
&  \Bigl( t^3 \,  
\left( 1 + t + 2\,{t^2} + {t^3} + 2\,{t^4} + {t^6} \right) \cr &
-\lambda\,  t \,  ( 1 + 3\,t + 6\,{t^2} + 9\,{t^3} + 
          11\,{t^4} + 11\,{t^5} \cr &  
+ 10\,{t^6} + 8\,{t^7} + 5\,{t^8} + 3\,{t^9} + 
          2\,{t^{10}} + {t^{11}} ) \cr & +
\lambda^4\,  t^3 \, ( 1 + t + 3\,{t^2} + 2\,{t^3} + 4\,{t^4} + 
2\,{t^5} + 4\,{t^6} 
\cr & 
 +2\,{t^7} + 3\,{t^8} + {t^9} + 2\,{t^{10}} + {t^{11}} + {t^{12}} + 
          {t^{14}} ) \cr & +
\lambda^2 \,  
( 1 + 2\,t + 7\,{t^2} + 10\,{t^3} + 18\,{t^4} + 19\,{t^5}  
        \cr & +  24\,{t^6}
 + 19\,{t^7} + 20\,{t^8} + 13\,{t^9} + 12\,{t^{10}} + 
          6\,{t^{11}}
 + 5\,{t^{12}} + 2\,{t^{13}} + 2\,{t^{14}} ) \cr & -
\lambda^3 \, t \,   
( 1 + 3\,t + 6\,{t^2} + 10\,{t^3} + 
          13\,{t^4} + 15\,{t^5} + 15\,{t^6}\cr &
 + 14\,{t^7} + 12\,{t^8} + 
          10\,{t^9} + 8\,{t^{10}} + 6\,{t^{11}} + 4\,{t^{12}}
 + 2\,{t^{13}} +2\,{t^{14}} + {t^{15}} ) \Bigr),
\cr}}
\eqn\fthreeone{
\eqalign{ f_{\tableau{3 1}}(t, \lambda)=&- 
{t^{-{9/2}} \over t^{1 \over 2} -t^{-{1\over 2}}} 
\left ( -1 + \lambda \right) ^2 \,
      {\lambda^2}\,\left( 1 + t \right) \,
      \left( \lambda - t \right) \, 
      \left( -1 + \lambda\,t \right) \cr &
      \Bigl( t^3 \,( 1 + t + {t^2} + {t^3})
+ \lambda^2 \,( 1 + t + {t^2} + {t^3} )^2 \,
         \left( 1 + t + 2\,{t^2} + {t^4} \right)\cr &
- \lambda \, t \,\left( 1 + 3\,t + 5\,{t^2} + 8\,{t^3} + 7\,{t^4} + 
           6\,{t^5} + 3\,{t^6} + 2\,{t^7} \right) \cr &
+ \lambda^4 \, t^2 \, 
\left( 1 + t + 3\,{t^2} + 2\,{t^3} + 3\,{t^4} + {t^5} + 2\,{t^6} + 
           {t^8} \right) \cr &
- \lambda^3 \,  t \,( 2 + 4\,t + 8\,{t^2} + 10\,{t^3} + 11\,{t^4} + 10\,{t^5} +  7\,{t^6} + 4\,{t^7} \cr &
 + 3\,{t^8} + {t^9} + {t^{10}} ) 
\Bigr), \cr}}
\eqn\fonethree{
\eqalign{
f_{\tableau{2 1 1}}(t, \lambda)=&{t^{-{19/2}} 
\over t^{1 \over 2} -t^{-{1\over 2}}} 
\left( -1 + \lambda \right) ^2 \,
      {\lambda^2}\,\left( 1 + t \right) \,
      \left( \lambda  - t \right) \,
      \left( -1 + \lambda\,t \right) \cr & 
\Bigl( {t^5}\,\left( 1 + t + {t^2} + {t^3} \right) 
+ \lambda^2 \, t\,{{\left( 1 + t + {t^2} + {t^3} \right) }^2}\,
        \left( 1 + 2\,{t^2} + {t^3} + {t^4} \right)\cr &
- \lambda \, t^3 \, 
\left( 2 + 3\,t + 6\,{t^2} + 7\,{t^3} + 
          8\,{t^4} + 5\,{t^5} + 3\,{t^6} + {t^7} \right) \cr & 
+\lambda^4 \, \left( t + 2\,{t^3} + {t^4} + 3\,{t^5} + 
          2\,{t^6} + 3\,{t^7} + {t^8} + {t^9} \right)\cr & 
- \lambda^3 \, 
( 1 + t + 3\,{t^2} + 4\,{t^3} + 7\,{t^4} + 
 10\,{t^5} + 11\,{t^6} + 10\,{t^7} \cr &
 + 8\,{t^8} + 4\,{t^9} + 2\,{t^{10}}) \Bigr),\cr}}
\eqn\ftwotwo{
\eqalign{
f_{\tableau{2 2}}(t, \lambda)=& -t^{-6} 
\left( -1 + \lambda \right) ^2\,
     {\lambda^2}\,\left( \lambda - t \right) \,
     \left( -1 + \lambda\,t \right) \,
     \left( 1 + t + {t^2} \right) \cr &
      \Bigl( 
{t^3} + {t^5} - \lambda \, t \,
         \left( 1 + t + {t^2} + {t^3} \right)^2 \cr &
- \lambda^3 \,t\,{{\left( 1 + t \right) }^2}\,
         \left( 1 + t + {t^2} + {t^3} + {t^4} \right) + 
\lambda^4 \,\left( t + 2\,{t^3} + {t^4} + 2\,{t^5} + 
           {t^7} \right)
  \cr &
+\lambda^2 \,
         \left( 2 + 2\,t + 6\,{t^2} + 6\,{t^3} + 9\,{t^4} + 6\,{t^5} + 
           6\,{t^6} + 2\,{t^7} + 2\,{t^8} \right)
 \Bigr), \cr} }
\eqn\ffouranti{
\eqalign{
f_{\tableau{1 1 1 1}}(t, \lambda)=&
-{ t^{-35/2} \over t^{1\over 2} -t^{-{1\over 2}}}
\left( -1 + \lambda \right) ^2\,
      {\lambda^2}\,
      \left(\lambda - t \right) \,
      \left( -1 + \lambda\,t \right)\cr & 
\Bigl( {t^8}\,\left( 1 + 2\,{t^2} + {t^3} + 2\,{t^4} + {t^5} + {t^6}
            \right)
\cr &-\lambda \,{t^5}\,
         ( 1 + 2\,t + 3\,{t^2} + 5\,{t^3} + 8\,{t^4} + 10\,{t^5} + 
           11\,{t^6} + 11\,{t^7} + 9\,{t^8}\cr & 
 + 6\,{t^9} + 3\,{t^{10}} + 
{t^{11}} )  + \lambda^4 \, ( 1 + {t^2} + {t^3} + 2\,{t^4} + {t^5} 
\cr &  + 3\,{t^6} + 2\,{t^7} + 
           4\,{t^8} + 2\,{t^9} + 4\,{t^{10}} + 2\,{t^{11}} + 3\,{t^{12}} + 
           {t^{13}} + {t^{14}} )  \cr & 
+ \lambda^2 \,{t^3}\,
         ( 2 + 2\,t + 5\,{t^2} + 6\,{t^3} + 12\,{t^4} + 13\,{t^5} + 
           20\,{t^6} + 19\,{t^7} + 24\,{t^8} \cr & 
 + 19\,{t^9} + 18\,{t^{10}} + 
           10\,{t^{11}} + 7\,{t^{12}} + 2\,{t^{13}} + {t^{14}} )  - 
       \lambda^3 \,t\,
         ( 1 + 2\,t + 2\,{t^2} + 4\,{t^3} \cr &
 + 6\,{t^4} + 8\,{t^5} + 
           10\,{t^6} + 12\,{t^7} + 14\,{t^8} + 15\,{t^9} + 15\,{t^{10}} + 
           13\,{t^{11}}\cr &
 + 10\,{t^{12}} + 6\,{t^{13}} + 3\,{t^{14}} + {t^{15}}
            )
\Bigr). \cr}}

\listrefs
\bye